\documentclass{article}

\usepackage{arxiv}

\usepackage[utf8]{inputenc} 
\usepackage[T1]{fontenc}    
\usepackage{hyperref}       
\usepackage{url}            
\usepackage{booktabs}       
\usepackage{amsfonts}       
\usepackage{nicefrac}       
\usepackage{microtype}      
\usepackage{graphicx}

\usepackage{soul}

\newcommand{\ignore}[1]{}

\newcommand{\arch}{ISER}

\newcommand{\recomp}{VRC} 
\newcommand{\rc}{RC}     

\newcommand{\redHL}[1]{\textcolor{red}{#1}}

\newcommand{\shrink}{Eliminate vertical white-space}
\newcommand{\vshrink}[1]{
  \ifdefined\shrink 
	\vspace{-#1cm}
  \else
	\vspace{0cm}
  \fi
}

\newcommand{\changed}[2]{\textcolor{red}{\sout{#1}}{\textcolor{blue}{#2}}}

\usepackage{subfig}
\newcommand{\squishlist}{
   \begin{list}{$\bullet$}
    { \setlength{\itemsep}{0pt}      \setlength{\parsep}{1pt}
      \setlength{\topsep}{1pt}       \setlength{\partopsep}{0pt}
      \setlength{\leftmargin}{1em} \setlength{\labelwidth}{0.8em}
      \setlength{\labelsep}{0.3em} } }

\newcommand{\squishend}{
    \end{list}  }

\title{On Value Recomputation to Accelerate Invisible Speculation}

\author{

Christos Sakalis \\
Uppsala University, Sweden \\
 \texttt{christos.sakalis@it.uu.se} \\
   \And
Zamshed I. Chowdhury \\
University of Minnesota, Twin Cities, USA\\
\texttt{chowh005@umn.edu}
\And
Shayne Wadle\\
University of Wisconsin, Madison, USA \\
\texttt{swadle@cs.wisc.edu}

\And
Ismail Akturk\\
University of Missouri, Columbia, USA\\
\texttt{akturki@missouri.edu}

\And
Alberto Ros \\
University of Murcia, Spain\\
\texttt{aros@ditec.um.es}

\And
Magnus Själander\\
Norwegian University of Science and Technology, Norway\\
\texttt{magnus.sjalander@ntnu.no}

\And
Stefanos Kaxiras \\
Uppsala University, Sweden\\
\texttt{stefanos.kaxiras@it.uu.se}
\And
Ulya R. Karpuzcu\\
University of Minnesota, Twin Cities, USA \\
\texttt{ukarpuzc@umn.edu}
}

\begin{document}
\maketitle

\begin{abstract}
Recent architectural approaches that address speculative side-channel attacks aim to prevent software from exposing the microarchitectural state changes
of transient execution.
The \emph{Delay-on-Miss} technique is one such approach, which
simply delays loads that miss in the L1 cache until they become non-speculative, resulting in no transient changes in the memory hierarchy.
However, this costs performance, prompting the use of value prediction (VP) to regain some of the delay.

However, the problem cannot be solved by simply introducing a new kind of speculation (value prediction).
Value-predicted loads have to be validated, which cannot be commenced until the load becomes non-speculative.
Thus, value-predicted loads occupy the same amount of precious core resources (e.g., reorder buffer entries) as Delay-on-Miss.
The end result is that VP only yields marginal benefits over Delay-on-Miss.

In this paper, our insight is that we can achieve the same goal as VP (increasing performance by providing the value of loads that miss) without incurring its negative side-effect (delaying the release of precious resources), 
if we can safely, non-speculatively, recompute a value in isolation (without being seen from the outside), so that we do not expose any information
by transferring such a value via the memory hierarchy.
\emph{Value Recomputation}, which trades computation for data transfer 
was previously proposed in an entirely different context: to reduce energy-expensive data transfers in the memory hierarchy. 
In this paper, we demonstrate the potential of value recomputation in relation to the Delay-on-Miss approach of hiding speculation, discuss the trade-offs,
and show that we can achieve the same level of security, reaching 93\% of the unsecured baseline performance (5\%
higher than Delay-on-miss), and exceeding (by 3\%) 
what even an oracular (100\% accuracy and coverage) value predictor could do.
\end{abstract}


\section{Introduction}
\label{sec:intro}

With the disclosure of Spectre~\cite{kocher_spectre_2018} and Meltdown~\cite{lipp_meltdown_2018} in early 2018, \emph{speculation}, one of the fundamental techniques for achieving high
performance, proved to be a significant security hole, leaving the door wide
open for side-channel attacks~\cite{bernstein2005cache,yarom_flush+_2014,liu15llc,irazoqui_cross_2016} to ``see'' protected data~\cite{kocher_spectre_2018,lipp_meltdown_2018}.
As far as the instruction set architecture (ISA) and the target program are
concerned, this type of information leakage
through
microarchitectural ($\mu$-architectural) state and structures is not illegal
because it does not violate the functional behavior of the program.
But \emph{speculative} side-channel attacks reveal secret information during \emph{misspeculations}, 
i.e., discarded execution that is not a part of the normal execution of a program.
The stealthy nature of a speculative side-channel attack is based on
\emph{microarchitectural} state being changed by misspeculation even when the \emph{architectural}
state is not.

\noindent \textbf{First response techniques: delay, hide\&replay, or cleanup?}
A number of techniques have already been proposed to prevent \emph{microarchitectural} state from leaking information during speculation, either by %
\emph{delaying} such effects~\cite{sakalis+:ISCA2019vp,weisse2019nda,yu_speculative:MICRO2019-STT, fustos+:DAC2019spectreguard}, %
\emph{hiding} them and making them re-appear for successful speculation (\emph{hide\&replay})~\cite{yan_invisispec:MICRO2018,sakalis+:CF2019ghost} or %
\emph{cleaning up} the changes when speculation fails~\cite{saileshwar2019cleanupspec}. %
Because these techniques were proposed for different threat models (i.e.,  responding to a different set of known or unknown threats), provide different protection for parts of the system that can leak secrets (e.g., caches, DRAM, core), and make different assumptions for what other parts of the system are protected (hence carry different costs), a direct comparison of all of them is, as of yet, not feasible. In this paper, without loss of generality, we focus on delay techniques and for convenience we adopt the threat model of the work of Sakalis et al., referred to as \emph{Delay-on-Miss  (DoM)}~\cite{sakalis+:ISCA2019vp}. 

\noindent \textbf{What problem are we solving?} Delay-on-Miss is the simple idea of delaying any speculative load that misses in the L1 cache until the earliest time when it becomes non-speculative. 
To recover some of the lost performance from delaying critical instructions (loads that miss) Sakalis et al. proposed to use \emph{value prediction} (\emph{VP}) for the delayed misses in hope of performing useful work for the delayed loads and their dependent instructions. In other words, the aim of VP is to increase instruction-level-parallelism (ILP) by executing dependent instructions using load-value prediction. 

The conundrum of this approach is the following: VP, as another form of speculation, forces predicted loads to be validated \emph{in-order} in the memory hierarchy, as each load remains speculative until all older loads have been performed non-speculatively. This means that the validation of these loads \emph{cannot have any memory-level-parallelism (MLP)}. 
\ignore{
\sout{This is the same problem many of the proposed techniques (e.g., ~\cite{yan_invisispec:MICRO2018,sakalis+:CF2019ghost,sakalis+:ISCA2019vp,weisse2019nda}}{\color{red} [CHECKME: is NDA affected?]}{\color{blue} -- CS: NDA prevents ILP/MLP by delaying instructions, it does not have validations, nor does it have anything to do with TSO. NDA does not have M-Shadows nor does it interact with coherence at all.}) \sout{are facing with a consistency model, such as TSO, that requires load order to be preserved.
However, VP imposes an order in the predicted loads, even when a consistency model, such as Release Consistency (\rc), does not require it, or even when the \emph{implementation} of a consistency model, such as TSO, allows \emph{non-speculative load-load reordering}~\cite{aros-isca17}.}
}
Thus, any possible gains in ILP from VP during speculation, could be compromised by the hindrance of MLP at validation~\cite{dom-tc2020}.

\noindent \textbf{A new perspective:} In this paper, we ask the question: Can we create ``secret'' values, \emph{invisible to an attacker}, for the delayed loads, without having to compromise MLP to validate them afterwards? Our key intuition is that the answer lies in \emph{value re-computation} (\emph{\recomp})
also known as \emph{Amnesic Computing}~\cite{amnesiac17}.
The idea is that recomputing a value on an L1 miss --- a value that otherwise would have been loaded from the memory hierarchy --- can replace the need to access the memory hierarchy.
This requires having a backward slice of producer instructions on a per (load) value basis, along with the necessary input operands to perform recomputation. By construction, slices do not contain any branch or memory references (be it a store or a load). 
Most importantly, recomputation is also not speculative by construction, hence prevents nested speculation (and negative side effects thereof).

\noindent \textbf{Our Contributions:} 
\squishlist
\item We propose to apply an unconventional idea, \emph{value recomputation} (previously proposed as a means to evade the cost of moving data in the memory hierarchy) to solve this problem. We devise a $\mu$-architectural framework for security-aware value recomputation, well fitted to the 
threat model at hand and show the synergy with Delay-on-Miss.
\item We evaluate the potential of \emph{value recomputation} in eliminating speculative metadata, which makes classic processors vulnerable to numerous threats, including but not limited to what is known so far.  
\squishend


\noindent \textbf{A summary of our results:} This is the first $\mu$-architectural proposal 
that has the potential of outperforming the (unsecured) baseline in terms of performance and energy-efficiency, reducing the performance overhead of Delay-on-Miss by $42\%$. In this paper, we provide a quantitative discussion on how to unlock this potential. Practically, we cover (known or yet to come) threats posed by speculative memory reads.
%

\ignore{
\begin{verbatim}
    1 Problem statement: 
        Spectre, Meltdown, ++?
    2 Summary of recent arch. solutions
        Making speculative data invisible
        2.1 Overhead & complexity per se
        2.2 Limitations
        2.3 Coverage
    3 Proposed solution: 
        Recompute speculative load values
        3.1 Overhead & complexity 
            Put into perspective
        3.2 Limitations 
        3.3 Coverage
        3.4 Putting it all together:
            * Orthogonal to which, 
                alternative to what, ... wrt 2
            * Open new vulnerabilities?
\end{verbatim}
}

\section{Background}
\label{sec:back}
\subsection{Speculative Shadows} 
Sakalis et al. introduced the concept of \textbf{\emph{Speculative Shadows}} to reason about the earliest time an instruction becomes non-speculative and is considered safe to execute regardless of its effects on $\mu$-architectural state~\cite{sakalis+:CF2019ghost,sakalis+:ISCA2019vp}.
Speculative shadows can be of the following types: \emph{E-Shadows} are cast by any instruction that can cause an \textbf{exception}; 
\emph{C-Shadows} are cast by \textbf{control instructions}, such as branches and jumps, when either the branch condition or the target address are unknown or have been predicted but not yet verified; \emph{D-Shadows} are cast by potential \textbf{data dependencies} through stores with unresolved addresses (read-after-write dependencies); \emph{M-shadows} are cast by \textbf{speculatively executed memory accesses} that may be caught violating the ordering rules of a memory model (e.g., total store order---TSO) and therefore may need to be squashed; and \emph{VP-shadows} are cast by \textbf{value-predicted loads}~\cite{sakalis+:ISCA2019vp}. To be more specific, shadows demarcate regions of speculative instructions. So far, attacks have been demonstrated under the E-~\cite{lipp_meltdown_2018}, C-~\cite{kocher_spectre_2018}, and D-Shadows~\cite{CVE-2018-3693} only, but we cannot exclude future attacks using the rest.

\subsection{Threat Model}
\label{sec:threat}

We target speculative side- or covert-channel attacks that utilize the memory hierarchy (caches, directories, and the main memory) as their side-channel. Non-speculative cache side-channel attacks, as well as attacks that use other side-channels (such as port contention) 
{are not covered by Delay-on-Miss and, although still possible,} are outside the scope of this work.
We make no assumptions as to where the attacker is located in relation to the victim (on the core) or if they share the same virtual memory address space or not.
{As a matter of fact the attacker and the victim can be the same process}, as in the Spectre v1 attack~\cite{kocher_spectre_2018}.
We assume that the attacker can execute arbitrary code or otherwise redirect the execution of running code arbitrarily.
How the attacker manages to execute or redirect such code is beyond the scope of this work. Instead of focusing on preventing the attacker from accessing data illegally, we focus on preventing the transmission of such data through a 
{cache or memory} side- or covert-channel.

In this work, we use the concept of speculative shadows to determine when a load is safe or not. Speculative shadows determine the earliest point at which an instruction is guaranteed to be committed and retired successfully. Other works, such as InvisiSpec~\cite{yan_invisispec:MICRO2018} and NDA~\cite{weisse2019nda}, make different assumptions based on the threat model. For example, InvisiSpec provides two different versions, one based on the initial Spectre attacks where only the equivalent of C-Shadows are considered as part of the threat model, and one based on protecting against all possible future attacks, utilizing all the shadows. Similarly, NDA provides different solutions if only C-Shadows are considered (strict/permissive data propagation), if D-Shadows should also be considered (bypass restriction), or if all shadows should be considered (load restriction). In this work we 
{follow the strictest approach and }assume that all shadows have the potential of being abused, as we cannot reasonably argue that any of them are not exploitable.

{\em To summarize, 
we cover any known or yet-to-be-discovered side-channel posed by a speculative memory read. We assume that all system components operate correctly.}

\subsection{Delay-on-Miss}
\label{sec:dom-vp}
The goal of Delay-on-Miss is to hide speculative changes in the memory hierarchy (including main memory). To achieve this, Delay-on-Miss delays speculative loads that miss in the L1 cache. Loads that hit in the L1 (and their dependent instructions) are allowed to execute speculatively as their effects (i.e., on the L1 replacement state) can be deferred to when the loads are cleared from any speculative shadow. 
The miss of a delayed load is allowed to be resolved in the memory hierarchy at the earliest point the load becomes non-speculative. An efficient mechanism to track shadows is proposed by Sakalis et al.~\cite{sakalis+:ISCA2019vp}. 

Under Delay-on-Miss, the vast majority of loads are executed speculatively (80+\% on average~\cite{sakalis+:ISCA2019vp}), which causes a notable fraction of the loads to be delayed. This takes up precious resources (i.e., entries in the instruction queue, the reorder buffer, and the load/store queue) and eventually stalls instructions from committing. 
The significant amount of speculation that is performed, results in each load being covered by several speculative shadows (five on average according to our simulations). This forces the majority of the loads to be executed serially, severely limiting MLP~\cite{clearing-shadows-pact2020, dom-tc2020}. Furthermore, removing any individual shadow (e.g., the C-Shadow) has a limited effect, as the load can be covered by another overlapping shadow~\cite{clearing-shadows-pact2020}.

\subsection{Delay-on-Miss and Value Prediction} 

The concept behind using value prediction (VP) with Delay-on-Miss is to speed up the delayed loads (and their dependent instructions) and regain some of the lost performance. However, VP---no mater how good we make it (even under 100\% coverage and accuracy)---gives only a limited benefit on top of Delay-on-Miss~\cite{dom-tc2020}. 
VP clearly cannot regain the lost performance because of the following:

VP cannot help much as it simply provides values early; however, the validation is still delayed until all shadows have been lifted. Thus, precious core resources are still occupied until the same point in time as simply delaying the load.
The only perceptible difference is a faster commit of pre-executed dependent instructions if the validation of a value-predicted load proves to be correct.

Furthermore, VP introduces a new speculative shadow, which is referred to as the \emph{VP-Shadow}. This new shadow is only lifted from younger loads when the validation of the VP is complete. Thus, preventing younger loads from validating in parallel and limits the MLP, which results in VP occupying precisous resources in the same manner as Delay-on-Miss.

\subsection{Value Recomputation}
\label{sec:recmp}

Due to imbalances in technology scaling,  the energy usage (and latency) of data transfers in the memory hierarchy can easily exceed the energy usage (and latency) of value recomputation~\cite{Horow}.
Value recomputation (\recomp) is proposed as a way to trade off data movement in the memory hierarchy for in-core computation to save energy~\cite{amnesiac17,taxonomy18}.
The basic idea is to swap slow and energy-hungry loads for recomputation of the respective data values.  
This is achieved by identifying a slice of producer instructions of the respective data values and executing them when the value is needed.
Each such slice forms a backward slice of execution, and strictly contains only arithmetic and logic instructions. 

\begin{figure*}
     \centering
     \subfloat[Slice example]{\includegraphics[width=0.3\textwidth]{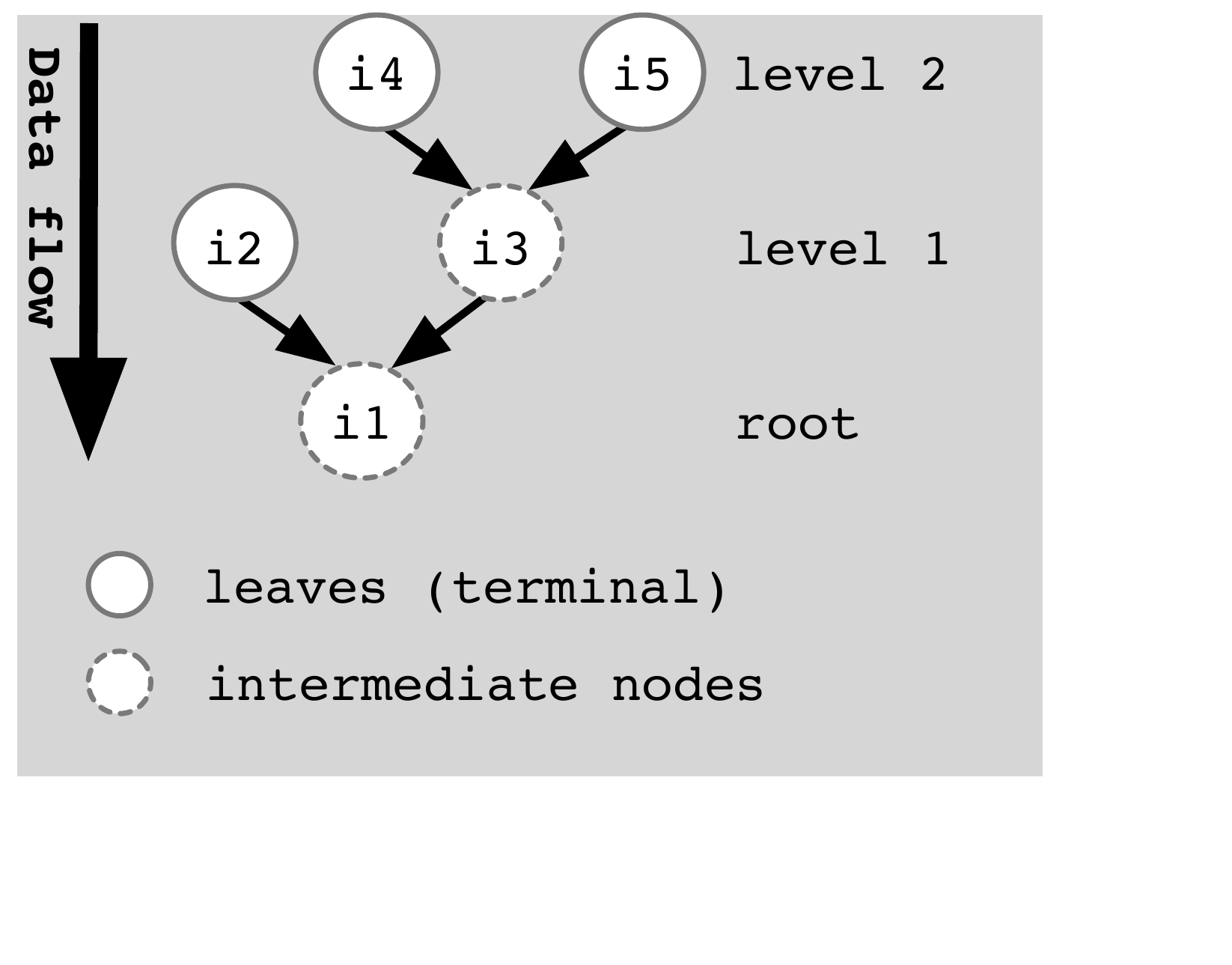}\label{fig:rt}}
     ~~~
     \subfloat[Execution semantics and  $\mu$-architecture]{\includegraphics[width=0.57\textwidth]{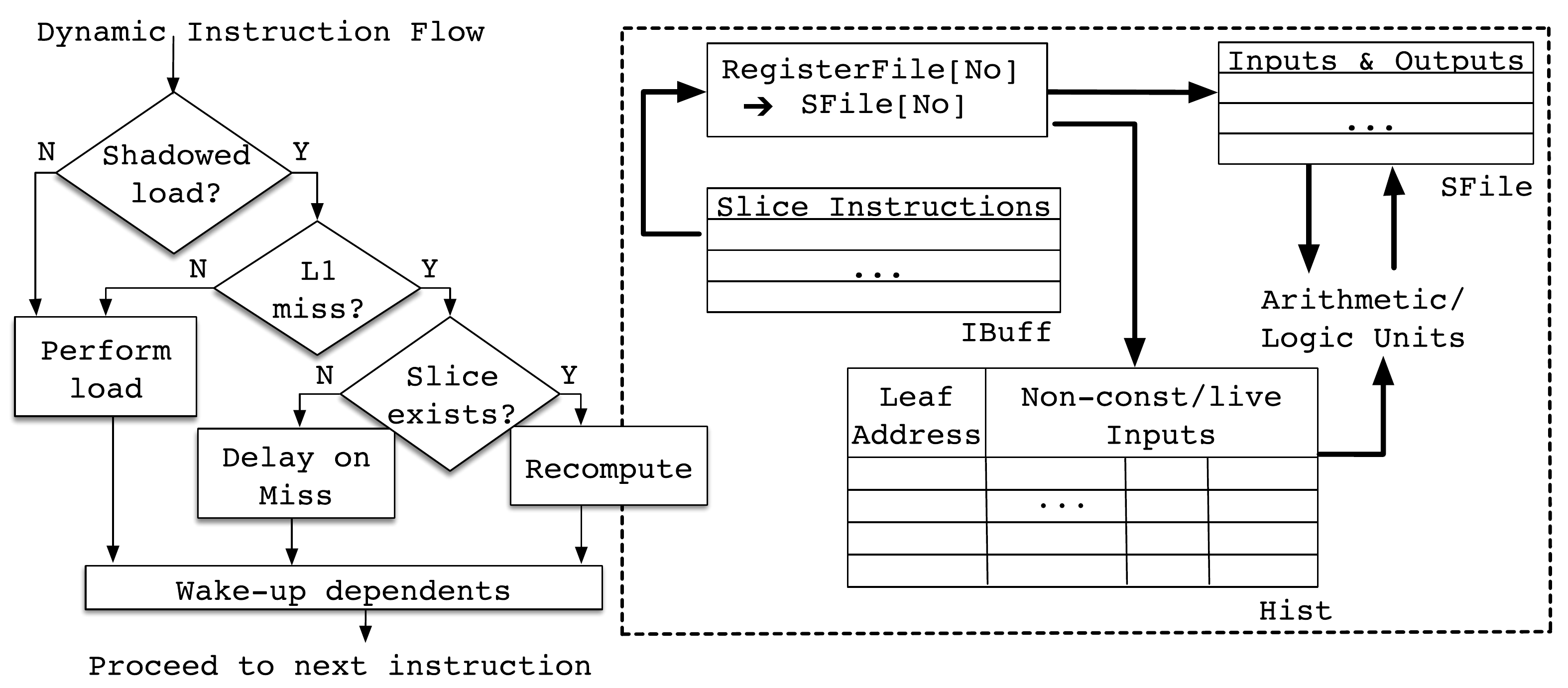}\label{fig:ctrl}}
     \caption{ (a) Backward slice; (b) \arch\ overview: All $\mu$-architectural buffers have  
an $\texttt{invalid}$ field per entry to
manage space (de)allocation.}
     \label{fig:overview}
\vshrink{.2}
\end{figure*}

As depicted in Figure~\ref{fig:rt}, each slice represents a data-dependency graph, where nodes correspond to producer instructions to be (re)executed. Data flows from the leaf nodes to the root. 
The root represents the producer of the store whose value will be recomputed when its corresponding (consumer) load is encountered, i.e., a load accessing the same memory location.
Nodes at level 1 are immediate producers of the (input operands of the) root, nodes at level 2 are producers of nodes at level 1, and so on and so forth. 
The nodes which do not have any producers are terminal instructions whose input operands must be available at the time of recomputation. If these input operands are read-only values to be loaded from memory (such as program inputs) or register values that will be overwritten, then buffering of these values are needed to enable recomputation of the load~\cite{amnesiac17}.

{\noindent \bf Premise:}
\recomp\ has the potential to render a more energy efficient (and faster) execution than servicing a miss in the memory hierarchy. At the same time, there is no need for MLP, since as opposed to VP, \emph{\recomp{} is not speculative} by itself and does not require any costly validation. 
A recomputed load can be committed as soon as all the shadows are lifted---this is in stark contrast to Delay-on-Miss/VP, which require a load/validation to be performed before commit.

\section{Invisible Speculative Execution through Recomputation (\arch)}
\label{sec:arch}
\ignore{
\begin{verbatim}
 * Overview
    High-level exec. semantics
    Coverage wrt threat model
 * Detailed (micro)architecture
    * Buffers
    * ISA extensions
 * Putting it all together
    [Flowcharts?]
    * Behavior under speculation
    * Behavior under non-speculative exec.
    * Relation to value prediction
* Overheads?
* Correctness guarantee?
* New vulnerabilities?
 \end{verbatim}
 }
 
We will next detail the mechanics of our novel approach, {\em Invisible Speculative Execution through (Value) Recomputation} (\arch). Due to space limitations, we will focus on how value recomputation can help eliminate the targeted threats (Section~\ref{sec:threat}). For a thorough discussion of value recomputation we refer the reader to~\cite{amnesiac17, taxonomy18}.
 
 \subsection{Execution Semantics}
 
 \arch{} only resorts to recomputation for regenerating values that otherwise would be read by a speculative load from the memory hierarchy, and only so, if the respective speculative load misses in the L1 cache. Recomputation takes place as long as a slice exists and the input operands to the slice instructions can be made readily available.
 
\ignore{
Not all of the input operands of leaf instructions of a slice can be
(re)generated by recomputation.  This may be the case if input operands
correspond to (i)~read-only values to be loaded from memory, such as program
inputs; or (ii)~register values that are lost, i.e., overwritten at the time of
recomputation. We will refer to such input operands as {\em non-recomputable}
inputs. For \recomp\ to work, non-recomputable inputs of slice-leaves
should not only be available at the anticipated time of recomputation, but also
be retrievable in an energy-efficient manner.
%
If non-recomputable inputs do not
reside in close physical proximity to the processor, the energy cost of their
retrieval may easily exceed the energy cost of memory accesses, rendering recomputation useless.    
Buffering can help in this case. No dedicated buffering is necessary if the leaf input operands
correspond to constants or live register values. 
}

While \arch\ shares basic $\mu$-architectural structures with Amnesiac~\cite{amnesiac17} to facilitate \recomp\ (such as dedicated buffers to prevent corruption of $\mu$-architectural state during recomputation), its
execution semantics are quite different when it comes to slice identification and triggering recomputation.
These stem from the defining difference in optimization targets: Amnesiac uses \recomp\ to maximize energy efficiency irrespective of security implications. \arch, on the other hand, uses \recomp\ to eliminate (already known or yet to be discovered) threats induced by speculative loads. In a nutshell, differences between Amnesiac and \arch\ expand along two axes:
 \begin{itemize}
     \item{\em What to recompute (slice identification):} As opposed to Amnesiac, \arch\ does not impose any direct constraint 
     to preserve energy efficiency,  as we are not after minimizing energy or latency per load. As long as a slice exists, and its inputs can be made readily available at the anticipated time of recomputation, \arch\ would consider it for recomputation. The only practical limitation on slice length may stem from storage overhead of $\mu$-architectural buffers in this case (Section~\ref{sec:iser-architecture}).
     \item {\em When to recompute:} \arch\ swaps speculative loads that miss in L1 for recomputation (i.e., with producer instructions of the respective value along a slice). Amnesiac on the other hand, triggers recomputation (irrespective of whether the load is speculative or not) only if it is more energy-efficient to do so. 
     \ignore{
     \item{\em When/how to commit a recomputed value to architectural state:}
     \textcolor{blue}{\\ HMS: {\recomp} is non-speculative and safe, they don't have to adhere to the shadows. It's the action of the load itself that has to be avoided, which it is, so the {\recomp} computed value can be treated as any other transient register, e.g., the result of an addition. Thus, I don't see any difference compared to Amnesiac. 
     U: Noted, and elimiated\\}
     In Amnesiac, a recomputed value is written into the destination register of the  load before execution resumes from where the corresponding load (which is swapped for recomputing instructions) has left. 
     \arch\ has to keep these values buffered until the speculation is resolved (i.e., all shadows are lifted).  
     \recomp\ cannot change architectural state under speculation, by definition. {However, just as speculative loads that \emph{hit} in the L1, recomputed loads pass their value (in a physical register) to dependent instructions and advance execution.} }
 \end{itemize}
 
 We continue with \arch{} design specifics, limitations, and side effects including coherence and consistency implications. 
 
 
\subsection{Slice Formation \& Annotation}
\label{sec:slice_formation}

Similar to Amnesiac, we rely on a compiler pass (backed up by profiling) to form and annotate slices, which 
 mainly constitutes
dependency analysis to
identify the producer instructions for each load.\footnote{Instead of a compiler,
the same job can be performed by \emph{dynamic binary instrumentation} at run time (albeit with probably inferior alias analysis but more dynamic information), rendering recompilation unnecessary in deployments where it is not an option.}
Slice creation is a \emph{best effort} under strict validity guarantees. Not being able to generate a recomputation slice for a load is not a security weakness under a security technique such as Delay-on-Miss, but simply a missed optimization opportunity.
 Although in this paper the slice formation is conservative, as we will see later, the requirement for strict guarantees of the recomputation validity can be relaxed (potentially increasing the coverage of recomputation, i.e., portion of load values that can be recomputed, and addressing coherence issues) if the appropriate architectural support is available. However, such extensions are outside the scope of this paper and will be fully evaluated in future work.

The slice formation pass builds the slice as a data dependency graph, where
the immediate producer of the value to be loaded resides at the root (Figure~\ref{fig:rt}).
As opposed to Amnesiac, the restriction to slice length comes from slice inputs or storage requirements (rather than the associated energy cost). 
If, during the traversal of data dependencies, we encounter other load instructions, we 
replace them recursively with the respective producer instructions. 
This recursive growth can continue until a store to the same address is encountered.
Loads and stores cannot be present 
in any slice by definition.

Once construction is complete, 
each slice gets embedded into the binary. 
Similar to Amnesiac, the special control
flow instruction $\texttt{RCMP}$ communicates recomputation opportunities to the runtime, which
semantically corresponds to an atomic bundle of a conditional branch + load (where no prediction is involved for the ``branch'' portion).
The runtime scheduler resolves the branching condition: if the respective load (while shadowed) misses in L1, $\texttt{RCMP}$ acts as a jump to the entry point (starting from the
terminal instructions) of the corresponding slice. Otherwise (i.e., the load is not shadowed or the shadowed load hits in L1), $\texttt{RCMP}$ acts as 
a conventional load.
All operands of the respective load and the starting address of its slice form the operands of the $\texttt{RCMP}$. 
An $\texttt{RTN}$ instruction (similar to a procedure return in nature) demarcates the end of each slice and returns the control to the instruction following the $\texttt{RCMP}$. Before the return takes place, the recomputed value is provided to the consumers of the respective load, in the same way as if the load was actually performed (i.e., by passing the value in a physical register). 

As explained by Akturk and Karpuzcu~\cite{amnesiac17}, recomputation is possible, even if the compiler cannot prove that all input operands of
terminal instructions correspond to immediate or live register values at the anticipated
time of recomputation, by keeping such input operands (e.g., overwritten register values) in a dedicated buffer.
For any operand of this sort, a $\texttt{REC}$
instruction is inserted directly after the instruction producing the value of the operand.
%
$\texttt{REC}$ takes as operands the destination register of the previous instruction and an
integer operand:
$\texttt{leaf-address}$, which points to the address of
the corresponding terminal instruction in the slice.  
$\texttt{REC}$ practically checkpoints the input
operand 
to a dedicated buffer. 
\subsection{\arch{} Architecture}
\label{sec:iser-architecture}

\arch{} implements the shadow tracking technique proposed by Sakalis et al.~\cite{sakalis+:ISCA2019vp}. %
The shadow tracking consists of a \textit{shadow buffer} (SB) that acts as a circular buffer similar to the reorder buffer (ROB). %
When a shadow casting instruction enters the ROB a new entry is allocated at the tail of the SB. %
Every load that enters the ROB checks the SB and if not empty, an entry is allocated in a \textit{release queue} that associates the load with the youngest entry in the SB (i.e., its tail). %
The load remains speculative as long as the head of the SB is marked as unresolved and not equal to the SB entry associated with the load. %
This mechanism performs a simple comparison between the head of the release queue and the head of the shadow buffer to identify when loads exit all their shadows,
thus, avoiding the need for costly content addressable memory~(CAM) searches. %

On top of this, as depicted in Figure~\ref{fig:ctrl}, \arch\ uses a few small buffers
that serve two main purposes:
(1) Keeping $\mu$-architectural state intact during recomputation; 
(2) Making slice instructions and operands available at the time of recomputation.

{\em The Scratch-File (SFile)} acts as a small physical register file during recomputation. Specifically, while recomputation is in progress, all data flows through the SFile. Thereby, \arch\ preserves $\mu$-architectural state during recomputation. No structure beyond the SFile is necessary in this case, as no memory access instruction is permitted in a slice.
%
%
\noindent {\em The Rename logic} translates (architectural) register references of slice instructions to SFile entries.
%
{To this end, \arch\ can re-use the
rename logic of a conventional out-of-order processor 
with the addition of a small  dedicated set of rename tables.
}

{\em  The Instruction Buffer (IBuff)}
caches slice instructions in order to avoid
unnecessary pressure on the instruction cache.
Each entry of IBuff corresponds to a recomputing instruction.
Fetch logic fills IBuff while 
IBuff feeds the rename logic.
%
For each slice  where the input operands of terminal instructions represent immediate or live
 register values, no additional buffering is necessary.
 Otherwise,  
\arch\ keeps the input operands (such as overwritten register values) for each terminal instruction in a dedicated buffer called {\em The History Table (Hist)}.
 The address ($\texttt{leaf-address}$) and (non-constant, non-live) input operands of a terminal
instruction constitute each Hist entry.

For \arch, an $\texttt{RCMP}$ always translates into \texttt{branch on L1 miss} for speculative loads. 
As shown in Figure~\ref{fig:ctrl}, for each $\texttt{RCMP}$ instruction encountered, \arch\ first checks whether the corresponding load is speculative, and if so, whether it misses in L1. %
Here we define an L1 miss as \textit{(i)} the cache block does not reside in the L1 cache and \textit{(ii)} there is no MSHR entry for that cache block. %
If an MSHR already exists it is then safe to take advantage of the existing MLP and service the load as soon as the older load is completed (i.e., the load that caused the MSHR to be allocated). %
\arch\ triggers recomputation for any shadowed load that misses in~L1. %
%
{An $\texttt{RCMP}$ instruction will always produce a value (either loaded or recomputed) so for each $\texttt{RCMP}$ a physical register is allocated by the conventional renaming mechanism. %
}

{
%
}
On a speculative L1 miss, \arch{} jumps to the entry point of the corresponding slice and starts fetching instructions. %
Inputs to a slice instruction can either come from 
\textit{(i)}~live register inputs, 
\textit{(ii)}~live values stored in Hist, or 
\textit{(iii)}~temporary values written to the SFile by an older slice instruction. %
Live registers are read directly from the physical register file using the conventional renaming tables. %
Architectural registers written by slice instructions are mapped to the SFile registers using the rename logic, similar to how conventional renaming would map them to the physical register file. %
Potential values stored in Hist are referenced using the address of the slice (leaf) instruction. %
%
Instructions are fetched until hitting $\texttt{RTN}$, which copies the produced value from the SFile 
to the physical destination register and wakes-up consumers of the recomputed value. %
The $\texttt{RCMP}$ instruction is then committed as any other instruction without further delays. 

\begin{figure}[t]
\centering
  \includegraphics[width=0.5\columnwidth]{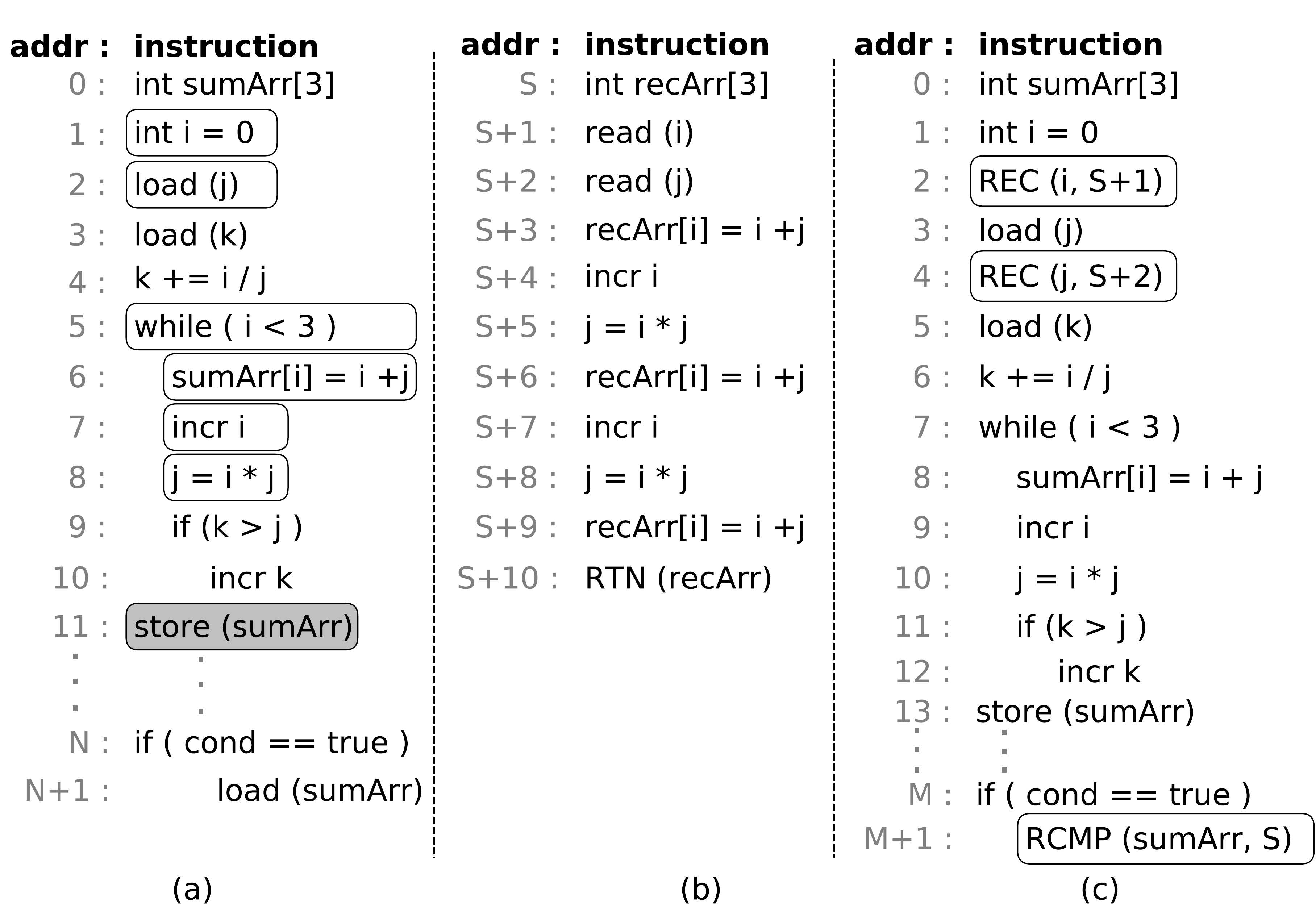}
  \caption{
  {Illustration of (a) slice identification; (b) slice generation; and (c) VRC-enabled code.} }
  \label{fig:rslice-running-example}
  \vshrink{.2}
\end{figure}

{Figure~\ref{fig:rslice-running-example} provides an illustrative example.}
{
Figure~\ref{fig:rslice-running-example}(a)
shows a pseudo-code
excerpt,
where we want to create a backward slice for the stored data \textit{sumArr}, which will later be (speculatively) loaded (line $N+1$). The instructions within boxes in Figure~\ref{fig:rslice-running-example}(a) are involved in the calculation of \textit{sumArr}, which are identified by the compiler (notice that the store instruction for \textit{sumArr} is not part of the slice
but informs us about
the memory address of the corresponding value).
Figure~\ref{fig:rslice-running-example}(b) shows the resulting backward slice (i.e., only the instructions involved in generating the value of \textit{sumArr}). In this illustration, we assume that inputs to the slice -- i and j -- are stored in Hist by  $\texttt{REC}$ instruction that associates inputs with leaf-instructions at addresses $S+1$ and $S+2$, respectively, as shown in Figure~\ref{fig:rslice-running-example}(c). Notice that the slice does not contain any control flow instruction, thus the while loop used in Figure~\ref{fig:rslice-running-example}(a) to generate values of \textit{sumArr} is unrolled in   Figure~\ref{fig:rslice-running-example}(b). Following the semantic explained earlier, $\texttt{RCMP}$ instruction at address $M+1$ in Figure~\ref{fig:rslice-running-example}(c) replaces the ordinary 
load instruction. Recall that $\texttt{RCMP}$ works as an ordinary load instruction if it hits in L1. However, if it misses L1, $\texttt{RCMP}$ jumps to entry point of the corresponding slice (which is at address $S$ in Figure~\ref{fig:rslice-running-example}(b)), and thereby avoids any  access to the lower levels of the memory hierarchy. After jumping to the slice entry point, inputs to the slice which were recorded earlier can be read from Hist (by instructions at addresses $S+1$ and $S+2$ in Figure~\ref{fig:rslice-running-example}(b)), and the desired output can be recalculated by fetching and executing instructions in the slice. Notice that \textit{recArr} is used as a temporary place holder for recomputed value (which is allocated in SFile), to keep the content of memory address of \textit{sumArr} intact (i.e., recomputation has no side effect/change in existing architectural state of the ongoing computation).  Finally, the intended value of \textit{sumArr} is recomputed and returned by the $\texttt{RTN}$ instruction (by copying the \textit{recArr} to destination of load instruction). Then, the control flow jumps back to the next instruction following $\texttt{RCMP}$ in Figure~\ref{fig:rslice-running-example}(c). The instructions contained in boxes in Figure~\ref{fig:rslice-running-example}(c) are extra instructions to be added into the binary to facilitate VRC.}

\subsection{Limitations \& Side Effects}
\label{sec:limitations}
 
\noindent {\em Overhead:} 
Latency or energy per recomputing instruction in a slice is not any different than the non-recomputing, conventional counterparts. The only difference is that \arch\ executes these instructions using a dedicated instruction and data supply rather than the conventional instruction cache and physical register file/data cache.

\noindent {\em Coverage:}
We cannot guarantee that all speculative loads missing in L1 have a corresponding slice. 
This may be due to complex producer-consumer chains, which cannot be expressed by a chain of arithmetic/logic instructions only, and/or slice inputs that cannot be guaranteed to be available during recomputation. 
Furthermore, some values are not produced by the application and are impossible to recompute, such as I/O.

\noindent {\em Locality:}
Any speculative load that misses in L1 and gets replaced with recomputation would never reach the memory hierarchy. As a result, subsequent memory requests to the same cache block become more likely to miss in the cache hierarchy, as well. This adverse effect can easily degrade performance, but recomputation targeting such new misses may be able to recover some of the lost performance. 
We will discuss this effect further in the evaluation (Section~\ref{sec:eval}).
 
\noindent {\em Exception Handling (during Recomputation):}
Exception handling during recomputation should be rare as it simply re-executes a previously seen slice of instructions with equivalent inputs. However, in case an exception would be raised we revert back to the Delay-on-Miss alternative and simply wait until all the shadows have been lifted (no longer speculative) and execute the load as normal.

\noindent
{{\em Pipeline Integration:}
The only negative impact may be due to potential increase in the pressure on execution units, as execution units are shared with the rest of the instructions. However, recomputing instructions along a slice (which form a dependency chain) are executed sequentially, one at a time. The impact would, therefore, be one additional instruction competing for the respective functional unit at a time. This can also be regarded as an opportunity to utilize the cycles (and functional units) that could be wasted otherwise due to stalled instructions waiting on delayed loads.
}


\ignore{
\begin{verbatim}
Stranger Things go here for the moment 
until we figure out what to do with them.
\end{verbatim}
\redHL{U: Will embed this into the previous section where we talk about practical implementation.}
}

\subsection{Architectural Support for Slice Coherence}
\label{sec:coherence}
ISER is based on the premise that we do not have to validate recomputed values: \emph{{\recomp} is not a speculation} (i.e., it is not a prediction). This is certainly the case for \emph{immutable} values that we can safely recompute instead of fetching them from the memory hierarchy. As long as the compiler guarantees via alias analysis that recomputed loads access \emph{immutable} values (from the time they were written by the corresponding producer), the approach is compatible with any consistency model and coherence protocol, simply because neither is needed to ensure correctness. We evaluate this case which, however, restricts {\recomp} coverage and limits the potential gains. 

Here, we sketch one approach on how to increase coverage by relaxing the restrictions on slice formation but the actual mechanisms are beyond the scope of this paper. 
Our aim is to show that there is significant untapped potential in this direction. 
In the evaluation we show the upper bound for such a potential approach with an oracle model. 

The central question is what happens if it is not possible to statically ascertain the immutability of a load's value. 
In other words, what happens for recomputed values that are considered as immutable but there is a possibility, however small, that they can change by some unknown store. 
We refer to such values as \emph{mostly-immutable}.\footnote{Naturally, we are not targeting \emph{mutable} values as successful {\recomp} would likely be much \emph{less} prevalent.}

For mostly-immutable values, we still want to maintain the essential property for our purposes, that {\recomp} is not a prediction that needs to be validated. Instead, what we want is to be able to make a simple binary decision: to recompute (if the value has not changed) or not (if the value has changed). In other words, we never validate {\recomp}, but we expect that a store would \emph{prevent} future recomputation of loads that access the same address. 
This  implies that we must \emph{track} any possible change of the data that could be accessed by recomputed loads. 

For single-threaded applications,
handling the recomputation of mostly-immutable values, implies a mechanism to match the thread's \emph{own} stores to the recomputed loads and invalidate the corresponding {\recomp} slices when such matches are found.\footnote{We assume, for the single-threaded case, that we would not recompute loads that touch I/O space that can be changed by a device without seeing any of our own stores modifying that space.}
To enable such a mechanism, the target address of the producer instruction 
is saved as a tag for the corresponding slice in the ISER structures. This tag can be matched by future stores on the same address, to invalidate the slice (and cancel recomputation) by invalidating, selectively or in bulk, ISER structures.
Since we expect this to be a rare occurrence (for what we choose to recompute), we can optimize for the case when it does not happen: Producer tags (store target addresses) can be encoded in signatures (Bloom filters) and 
if a future store hits in a signature, ISER structures and signatures are reset in bulk and need to be repopulated anew.

For multithreaded-applications, this matching and invalidation of recomputation slices should be expanded to include stores from other threads besides the thread's own stores. This requires an additional ``coherence'' mechanism to detect remote writes even when there is no copy of the relevant cacheline in the local cache. 
A solution can be based on an approach that serves a similar purpose: detecting remote writes in the absence of cached copies.

Specifically, the \emph{Callback} concept, introduced by Ros and Kaxiras~\cite{ros2015callback} can serve as the substrate on which to build a solution. A callback simply says ``notify me if someone writes on this address'' and it does not need cached copies that invite invalidations. Callback was introduced for synchronization, as an explicit \emph{request} for an invalidation in the absence of coherence invalidations (or more broadly absence of sharing). Callback can be generalized to perform a similar role in our situation with regards to detecting changes on what we would otherwise consider immutable values.
Similarly to the single-threaded case, tracked addresses can be encoded in signatures for efficient matching.
Security implications of using callbacks (such as perhaps new side-channels enabled by the callback directories~\cite{ros2016racer}) must also be addressed in the same way as in the work of Yan et al., SecDir~\cite{yan2019secdir}.

To conclude, we argue that {\recomp} slices can be made \emph{coherent} by explicitly detecting changes to what we would consider immutable values. Techniques for explicitly detecting writes without invalidations have been proposed in prior work~\cite{ros2016racer,ros2015callback} and their adaptation to our purposes is feasible.

\subsection{Impact on Consistency}
\label{sec:consistency}
While the coherence approaches sketched above enable us to \emph{explicitly detect} changes in mostly-immutable values and \emph{invalidate} the corresponding {\recomp} slice, here, we discuss the order that this would need to happen in relation to the consistency model of the baseline architecture.
We use total store order (TSO) and release consistency ({\rc}) as our prime examples but our reasoning can be generalized to other consistency models. 
We use the term \emph{callback invalidation} to distinguish from the normal coherence invalidation, which may not be available when we have no cached copy of the corresponding data.
The question here is, once a change is detected to a value that we are capable of recomputing, when exactly is {\recomp} canceled? 

If {\recomp} occurs well in advance of the callback invalidation it is safe in any consistency model such as TSO or {\rc}. By ``well in advance'' we mean that the recomputed load is retired from the reorder buffer. In this case, it is as if the corresponding load has seen the \emph{old} value, well in advance of the change in the value. Once the callback invalidation reaches the core, there will be no further {\recomp} of that load.
Thus, we only need to clarify what happens when a callback invalidation and the corresponding {\recomp} occur in a critical window when consistency rules could be violated.

In {\rc}, {\recomp} is safe between memory fences. (RC, allows both loads and stores to be reordered, unless otherwise enforced by memory fences.) Callback invalidations received before an acquire memory fence must take hold and cancel {\recomp} before crossing the fence.

In TSO, load--load reordering is not allowed to be observed. A recomputed load is considered \emph{performed} as we consider it equivalent to accessing the actual data. In a \emph{speculative} implementation of TSO,  a recomputed load would be speculative with respect to an older load that is not performed. In other words, a recomputed load can be  in the M-Shadow of one or more older loads.
A callback invalidation reaching the core while a recomputed load is still under an M-Shadow (e.g., one or more older loads are still not-performed) should squash the recomputed load (and its dependents) and cancel further {\recomp}. 


To conclude, we argue that {\recomp} is compatible with both TSO and RC by observing a correct ordering between callback invalidations and {\recomp}.

%
%
%
%
%
%
%
%

\subsection{Recomputation Security}
\label{sec:recmp-sec}

ISER is based on slice formation, replacement of corresponding loads with $\texttt{RCMP}$ instructions, and checkpointing of input operands with $\texttt{REC}$ instructions. The question here is what happens if any part or all of the ISER infrastructure can be abused by an adversary. This is of course equivalent to hijacking the compiler, or dynamic instrumentation (or even the binary of an application where the same security risks would apply).
However, even under such assumptions, \emph{ISER still cannot leak information speculatively}, which is the main goal of our work.

To see this, assume that the compiler are compromised. Attackers can make them do anything they want. We are still safe with respect to leaking information via speculative side-channel attacks because of the following reasons:

\begin{enumerate}

    \item {\recomp} itself cannot be used to construct a speculative side-channel in the memory hierarchy because it does not perform any memory accesses at all.
    
    \item {\recomp} is only used if the load is already under a speculative shadow. Even if {\recomp} recomputes a secret value, all future loads will be restricted under Delay-on-Miss.
    
\end{enumerate}

{To expand on (2), {\recomp} only starts if the $\texttt{RCMP}$ is under a speculative shadow. While {\recomp} has access to input operands that may hold secrets, the recomputation slice cannot perform any memory accesses to leak those secrets and the only way would be to pass the secret value to another (younger) load, which \emph{will also be speculative.} Delay-on-Miss guarantees that the younger load cannot have any visible side-effects, preventing any information leakage.} Essentially, {\recomp} maintains the Delay-on-Miss invariant that only non-speculative loads are allowed to cause side-effects in the memory hierarchy. 
Therefore, we conclude that {\recomp} is \emph{safe} from \emph{{cache and memory} speculative side-channel attacks}, no matter how compromised the compiler, dynamic instrumentation, or the binary is.

{In addition, the {\recomp} structures are local to the core and cannot be observed by another core. While under speculation, the only changes allowed are ones that cannot be observed from the outside, such as writes to the SFile. Any other changes (e.g., to the IBuff and Hist) are buffered or squashed, i.e., they are only updated once the instruction causing the change is no longer speculative. Furthermore, if SMT is present, the {\recomp} structures can be partitioned where necessary, to avoid contention attacks between SMT threads. It should be mentioned that if SMT is present, since the slices use the functional units (FUs) of the core, it is possible to perform an FU-contention attack. However, such attacks are outside the scope of Delay-on-Miss and this work, and are possible with or without {\recomp}. Thus, {\em {\recomp} does not open up any new attack opportunities under our current threat model.}}
Note also that disabling SMT has been recommended by vendors (e.g., Microsoft~\cite{ms_defences}) as a measure against several attacks.


\section{Evaluation Setup}
\label{sec:setup}
\ignore{
\begin{verbatim}
    * Benchmarks 
        * Justification for non-parallel?
    * Simulation framework 
        * Microarchitecture for recomputation
            * Recomputation buffers
                Size, configuration, ...
            * Model for potential ISA extensions
        * Energy model
\end{verbatim}
}

\begin{table}
  \centering
  \caption{The simulated system parameters.}
  \label{table:sim-params}
  \small
  \begin{tabular}{ l | l }
    \hline
    Parameter & Value\\
    \hline
    Technology node                           & 22nm \\
    Processor type                            & out-of-order x86 CPU \\
    Processor frequency                       & 3.4GHz \\
    Issue / Execute / Commit width            & 8 \\
    Cache line size                           & 64 bytes \\
    L1 private cache size                     & 32KiB, 8-way \\
    L1 access latency                         & 2 cycles \\
    L2 shared cache size                      & 1MiB, 16-way \\
    L2 access latency                         & 20 cycles \\
    Value predictor                           & VTAGE \\
    Value predictor size                      & 13 comp.s $\times$ 128 entries \\
    \hline
  \end{tabular}
\end{table}

We use a Pin-based tool~\cite{pin} to identify and annotate recomputation slices. For practical reasons, we limit the maximum slice size during construction to 100 instructions (which represents a loose upper bound in practice).
The annotated slices, together with the original binary, are fed to the gem5~\cite{binkert_gem5:CANEWS2011} simulator where the shadows, Delay-on-Miss, and VP have been implemented as described in the Delay-on-Miss work by Sakalis et al.~\cite{sakalis+:ISCA2019vp}. 
In gem5, we begin with fast-forwarding through the first one billion instructions of the application and then simulate in detail for another billion. 
We use McPAT~\cite{li_mcpat:MICRO2009} with CACTI~\cite{li_CACTI:ICCAD2011}, as well as the dynamic DRAM energy provided by gem5, to calculate the energy breakdown of the system. 
The configuration used for simulations are shown in~\autoref{table:sim-params}.
We evaluate the following versions:

\begin{list}{\labelitemi}{\leftmargin=2.5em}
    \item[\bf Baseline:] An unsecured out-of-order CPU.

    \item[\bf DoM:] Delay-on-Miss without any value prediction or recomputation. This is considered as the \emph{secure} baseline.
    
    \item[\bf VP:] DoM with an added VTAGE value predictor. 
    
    \item[\bf VRC:] DoM with the added value recomputation. This is the solution we are proposing. This does not include callbacks, only immutable values are recomputed.
    
    \item[\bf VRC (2 cycles):] Same as the VRC version but we have \emph{artificially} limited the latency of every slice to at most two cycles. We have also limited the number of instructions needed for the recomputation accordingly. As all VP versions take at most 2-cycles per prediction in our implementation, this VRC version enables iso-performance comparison with VP variants. 
    
    \item[\bf Oracle VP:] Same as the VP version but with an oracle predictor capable of predicting correctly $100\%$ of all speculative L1 misses. Even though the predictor is perfect, its results are still being validated once the loads have been unshadowed.
    
    \item[\bf Oracle {\recomp}:] Same as the VRC (2 cycles) version but with an oracle compiler capable of recomputing $100\%$ of all speculative L1 misses.
    Note that this is the Oracle in regards to {\recomp} coverage, not performance. We discuss the implications of recomputing all speculative L1 misses in the evaluation, \autoref{sec:eval}.
\end{list}

For the sake of brevity, the last three versions are only shown in the performance (IPC) results and are excluded from the rest of the figures.
%

We evaluate all these different versions using the SPEC2006 benchmark suite~\cite{spec:cpu06}, with the reference inputs, as in previous work~\cite{sakalis+:ISCA2019vp}.
{For one of the benchmarks, \texttt{GemsFDTD}, none of the techniques we tried produced any improvement. \texttt{GemsFDTD} is a floating point benchmark that is dominated by overlapping C-Shadows.
It achieves only about 20\% of the baseline performance with DoM (also corroborated by Sakalis et al.}~\cite{sakalis+:ISCA2019vp}).
{
In our work, we were unable to achieve any improvement with either VP} 
{or} {\recomp} 
{because of near-zero coverage. 
In contrast, it shows an impressive $3.5\times$ ($350\%$) improvement with an oracle }{\recomp} 
{($100\%$ coverage)---however, this may be impractical to attain. 
Energy results follow the same pattern, either showing high energy consumption ($3\times$ of the baseline) with all the techniques we tried or $56\%$ lower than the baseline with the} {\recomp} 
{oracle.
We surmise that \texttt{GemsFDTD} performs badly, in general, under any ``delay'' technique (including NDA}~\cite{weisse2019nda} 
{and STT}~\cite{yu_speculative:MICRO2019-STT}
{). 
Unfortunately, it is not included in these works to allow for comparisons. 
Because \texttt{GemsFDTD} represents such a special case for delay techniques we believe that further work is required to specifically address its shortcomings. 
For these reasons, we point out its idiosyncrasy here, instead of discussing it with the rest of the benchmarks.
}

\section{Evaluation}
\label{sec:eval}
\ignore{
\begin{verbatim}
  * Impact on performance (IPC)
  * Impact on energy
  * Coverage
  * Deep Dive
    * Cache characterization 
        Locality impact?
    * Slice statistics 
        Instruction mix, length, ...
    * Recomputation overhead?
    * Sensitivity analysis?
    * ...
 \end{verbatim}
 }

\ignore{
\begin{figure}[t]
  \includegraphics[width=\columnwidth]{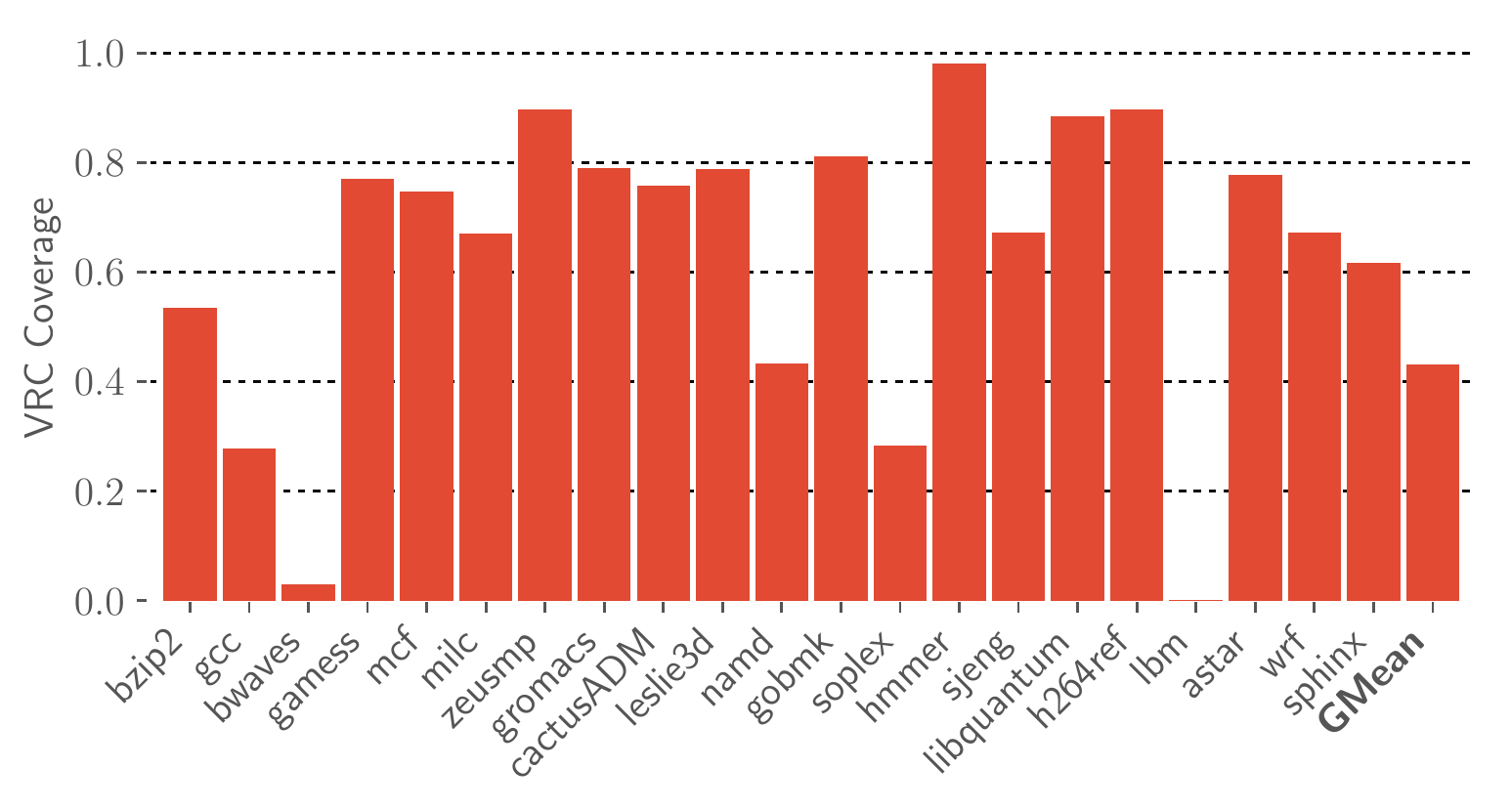}
  \caption{The coverage of the \recomp, i.e., the ratio of shadowed L1 misses that can be recomputed instead of being delayed.}
  \label{fig:rc-coverage}
\end{figure}

\begin{figure}[t]
  \includegraphics[width=.7\columnwidth]{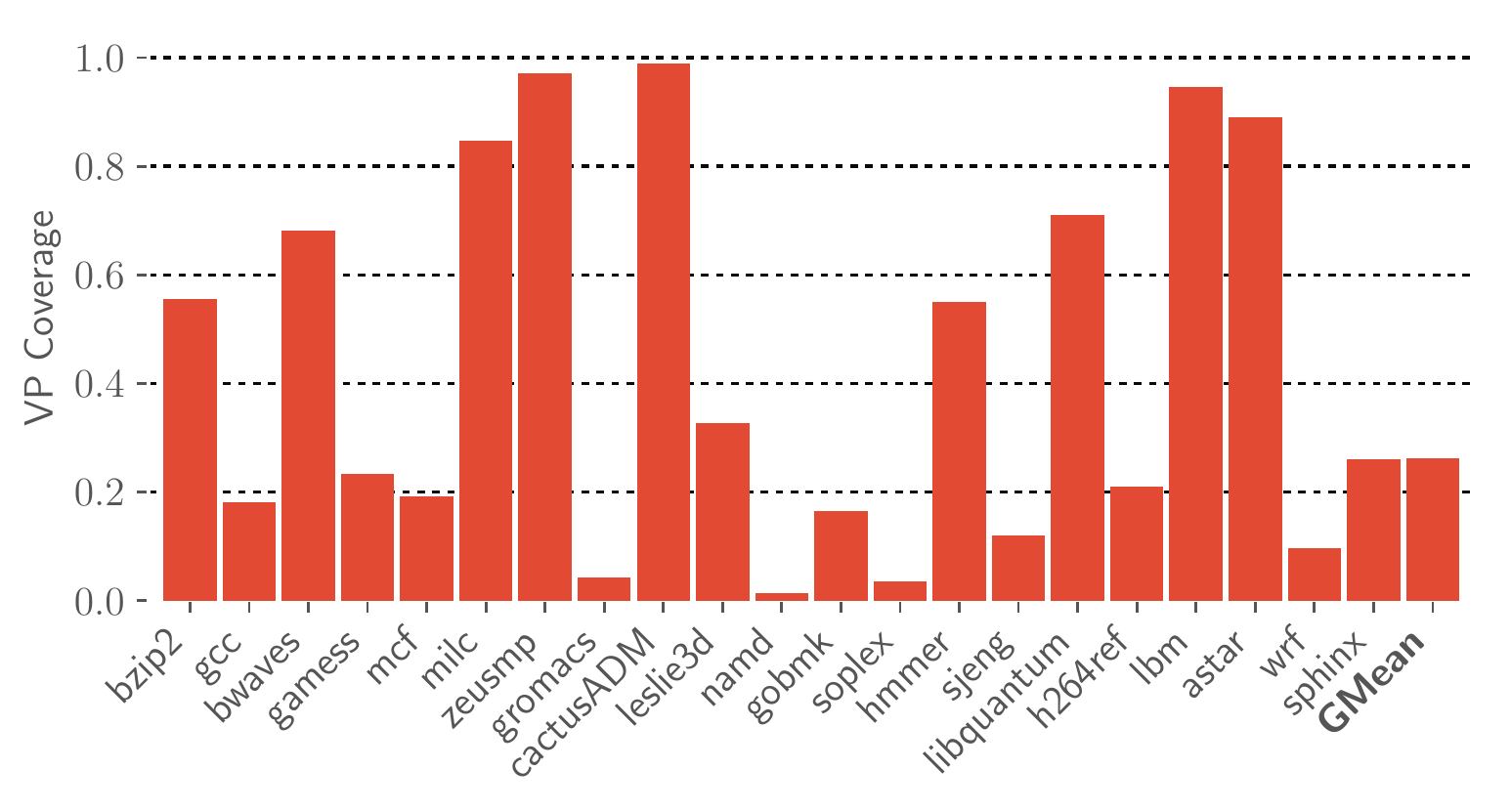}
  \caption{The coverage of the VP, i.e., the ratio of shadowed L1 misses that can be predicted instead of being delayed.}
  \label{fig:vp-coverage}
\end{figure}
}

\begin{figure}[t]
\vspace{-2ex}
\centering
  \includegraphics[width=.7\columnwidth]{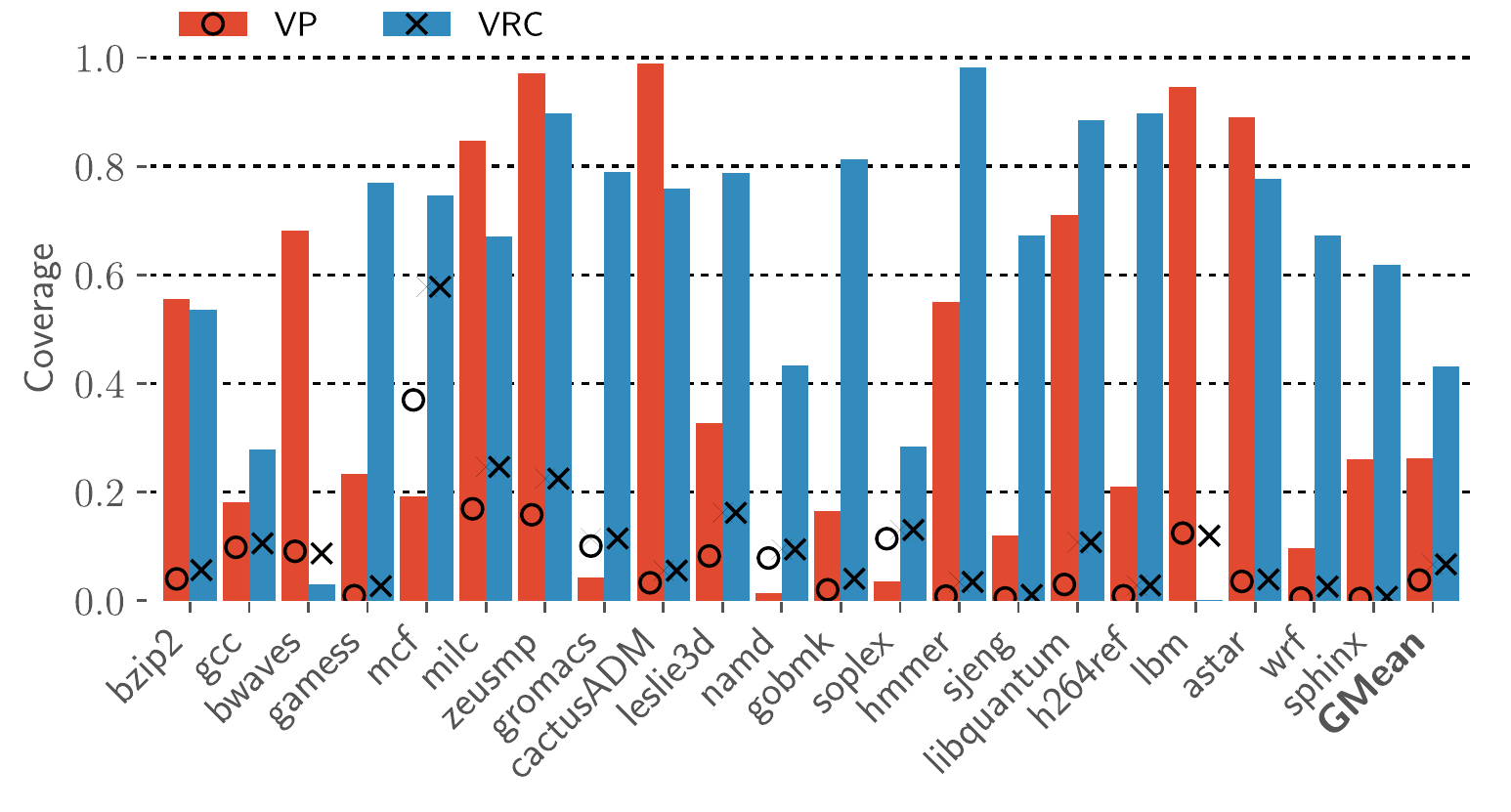}
  \caption{The coverage of VP and \recomp, i.e., the ratio of shadowed L1 misses that can be predicted or recomputed instead of being delayed (bars). Also depicted on the same plot is the L1 miss ratio for both versions (circles/crosses).}
  \label{fig:coverage}
\end{figure}

\begin{figure}[t]
\vspace{-2ex}
\centering
  \includegraphics[width=.7\columnwidth]{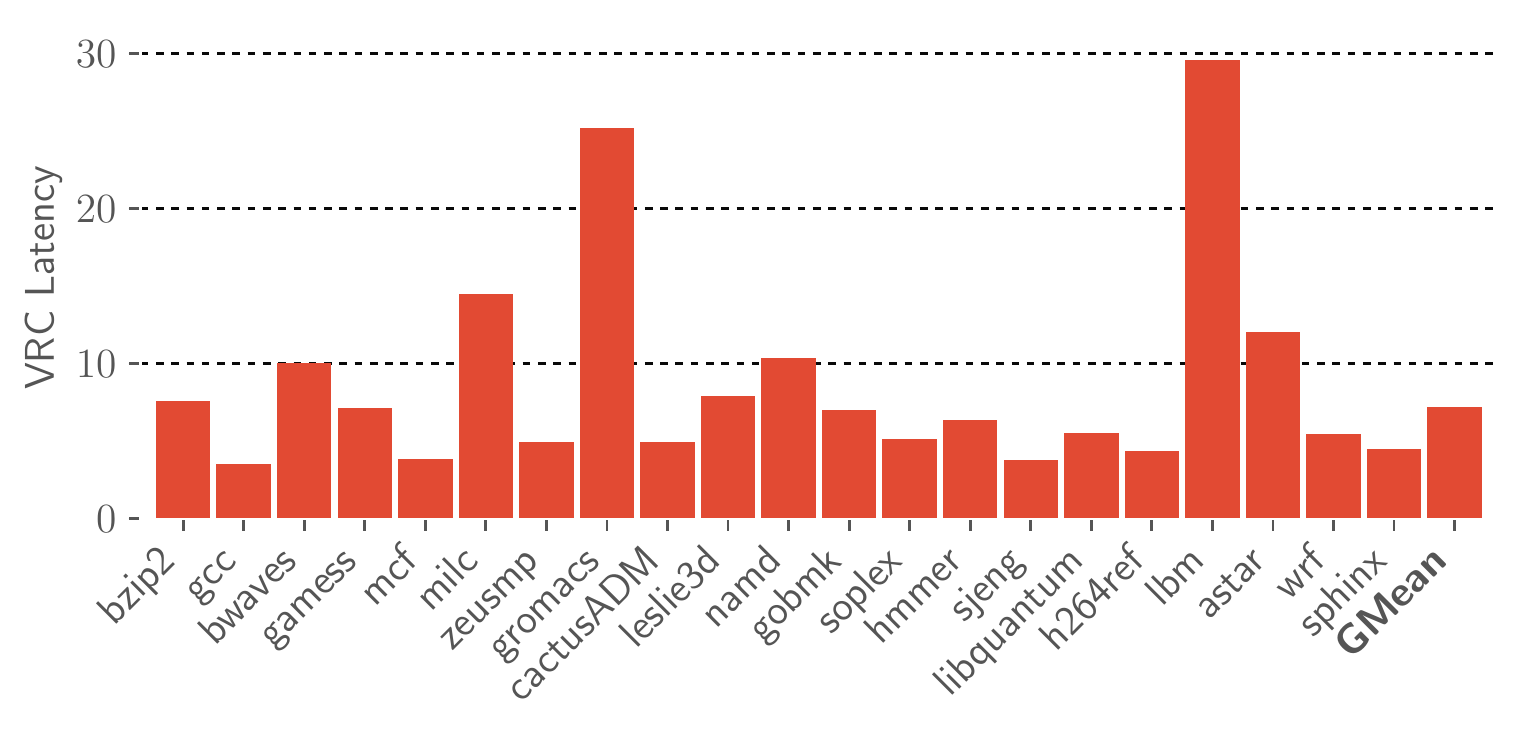}
  \caption{The mean latency for recomputing a shadowed L1 miss.}
  \label{fig:rc-latency}
\end{figure}

\begin{figure*}[t]
\vspace{-2ex}
  \includegraphics[width=\textwidth]{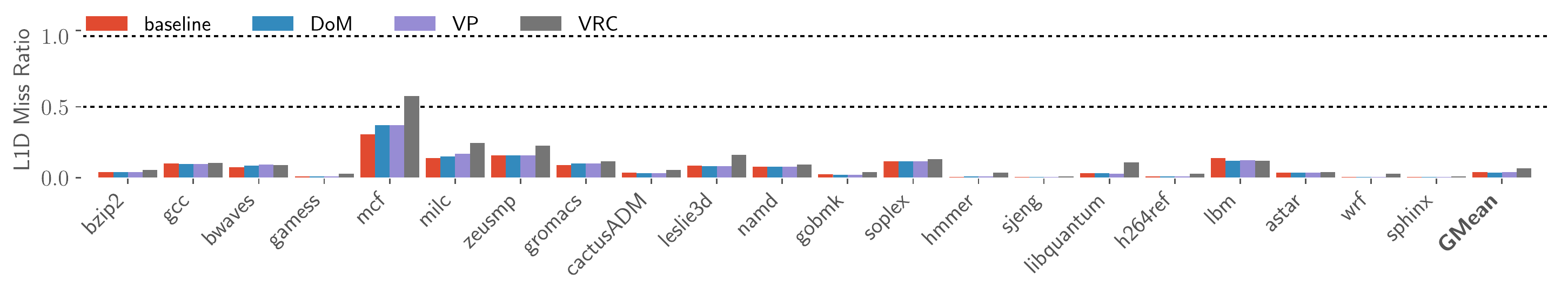}
  \caption{L1D miss ratio for Delay-on-Miss with VR and \recomp.}
  \label{fig:l1d_misses}
\end{figure*}

\begin{figure*}[t]
  \vspace{-2ex}
  \includegraphics[width=\textwidth]{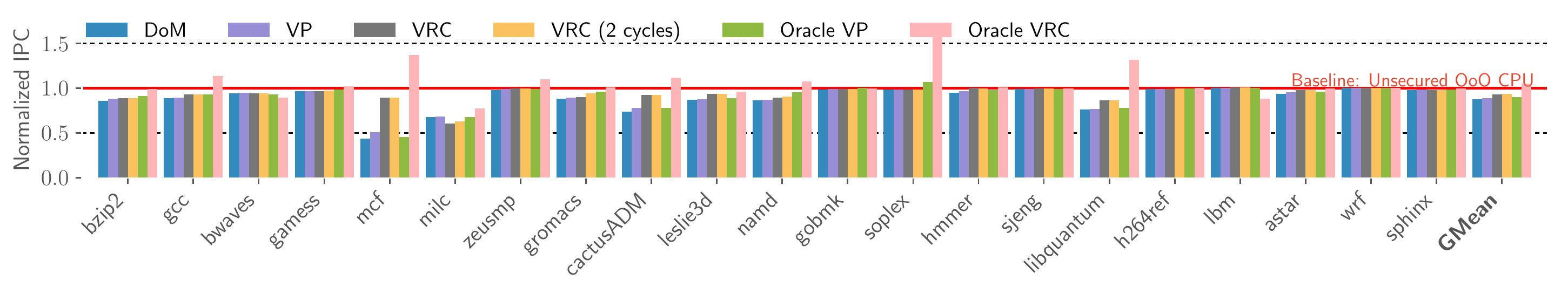}
  \caption{Performance (IPC -- higher is better) normalized to an unsecured OoO baseline.}
  \label{fig:ipc}
  \vspace{-2ex}
\end{figure*}

\ignore{
\begin{figure*}[t]
\vspace{-2ex}
  \includegraphics[width=\textwidth]{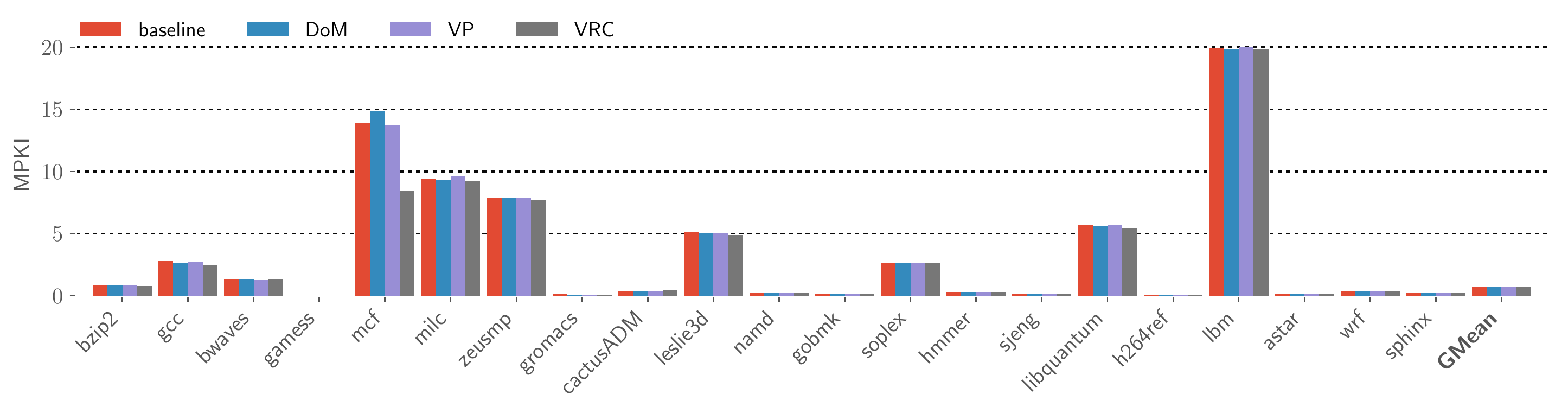}
  \caption{LLC misses per 1000 instructions (MPKI).}
  \label{fig:mpki}
\end{figure*}
}

\begin{figure*}[t]
  \includegraphics[width=\textwidth]{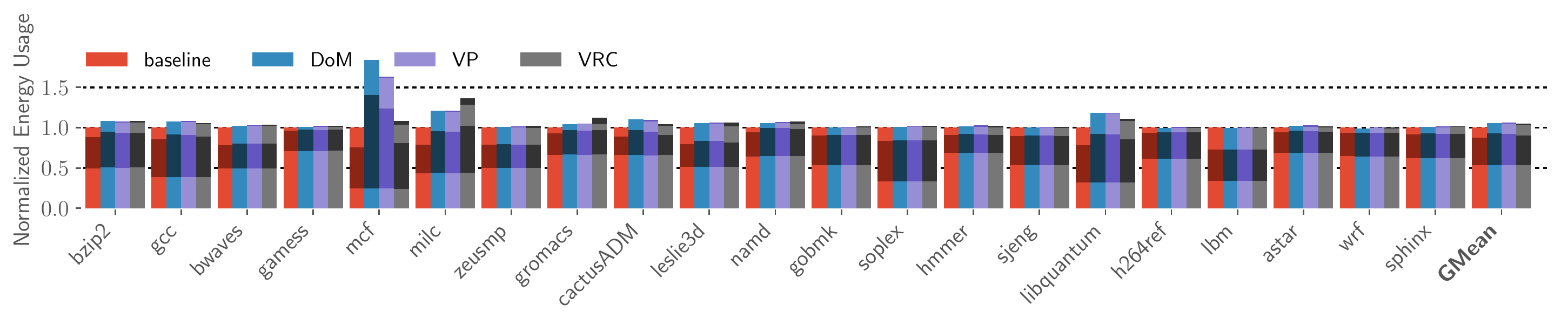}
  \caption{Energy usage, where each bar consists of four parts (from bottom up): The bottom, light colored part is the dynamic energy of the CPU, the middle, dark colored one is the static energy of the CPU, the middle light part is the DRAM energy, including refresh and power-down energy, and the top dark part is the overhead of VP and VRC, both static and dynamic.
  }
  \label{fig:energy}
\end{figure*}


\subsection{Recomputation Coverage}
\label{sec:coverage}

The coverage for the {\recomp} can be seen in \autoref{fig:coverage}, together with the VP coverage. We can immediately observe that, on average, VRC has higher coverage than VP, at $43\%$ of all speculative L1 misses vs. $26\%$ with the VP. A notable example is \texttt{mcf}, which is one of the worst performing benchmarks with DoM (Section~\ref{sec:performance}). On the other hand, \texttt{lbm} is a counter-example, where we have almost zero VRC coverage. This, however, does not affect the performance negatively, as \texttt{lbm} does not suffer from any performance penalties even with the plain DoM.

In the same figure, we have also superimposed the cache miss ratio for both versions. We only predict or recompute L1 misses, so the miss ratio is needed in conjunction with the coverage to infer the percentage of loads in the application that are being predicted or recomputed. More detailed L1D miss data can be found in \autoref{fig:l1d_misses}. Note how, as discussed in Section~\ref{sec:limitations}, \recomp{} increases the miss ratio.

With VP, all loads that can be predicted are predicted in the same amount of time (two cycles in our setup), but the same is not true for the VRC, where the latency depends on the slice length and the instructions it contains. In \autoref{fig:rc-latency} we can see the mean recomputation latency for each benchmark, as well as the overall mean. In all cases, VRC requires more cycles than VP to recompute a value, with a mean of seven cycles per slice. However, as we will see in Section~\ref{sec:performance}, this does not impact the performance significantly.

\subsection{Performance}
\label{sec:performance}

\autoref{fig:ipc} contains the number of committed instructions per cycle, normalized to the unsecured baseline processor. 
Delay-on-Miss without VP or \recomp, which is our \emph{secure} baseline, performs at $88\%$ of the unsecured baseline, similar to the results reported by Sakalis et al.~\cite{sakalis+:ISCA2019vp}. 
The benchmarks that incur the biggest hit in performance are \texttt{mcf} (at $44\%$ of the baseline), followed by \texttt{milc} ($68\%$), \texttt{cactusADM} ($74\%$) and \texttt{libquantum} ($76\%$).
Out of these benchmarks, three (\texttt{mcf}, \texttt{milc}, and \texttt{libquantum}) have high LLC MPKI,
but that in itself is not the only factor, as other benchmarks (e.g., \texttt{lbm}) also have a high MPKI. 
Instead, the cost of Delay-on-Miss also depends on the amount of MLP that the benchmarks exhibit; the more MLP that is taken advantage of in the baseline, the higher the performance loss. 

If VP is introduced, then the performance is similar, at $89\%$ of the unsecured baseline.
This result contradicts the results given by Sakalis et al.~\cite{sakalis+:ISCA2019vp}, where the VP gives a significant performance advantage\footnote{We contacted the authors and verified that our results are indeed valid.}. 
The reason that VP does not offer a significant advantage is because VP itself is speculative: When a value is predicted it still needs to be validated at a later point. 
By predicting the value, a small amount of parallelism (ILP) can be exploited during execution, but the slow L1 misses still need to be satisfied for the validation. 
Due to the high number of speculative shadows, validations become serialized and are not able to take advantage of any MLP that might be found in the application. 
In essence, the VP pushes the cost of delaying speculative loads from the execution stage to the validation stage, but it does not eliminate it. 
This can be seen in the Oracle VP results, where even $100\%$ prediction rate (i.e., all shadowed L1 misses are successfully predicted) only leads to a marginal performance improvement of one percentage point.

The same is not true for \recomp, as once a value has been recomputed, it does not need to be validated, meaning that the cost for delaying a long latency miss is eliminated and no  serialization is enforced.
While \recomp{} does not increase the amount of MLP that can be taken advantage of, it does eliminate some of the need for it.
Overall, \recomp{} performs at $93\%$ of the unsecured baseline, decreasing the performance cost of Delay-on-Miss by more than one third (specifically, by $42\%$).
The benchmark with the most dramatic performance increase is \texttt{mcf}, which is the worst performing benchmark for Delay-on-Miss. 
\recomp{} improves the performance from $44\%$ to $90\%$, reducing the performance cost to one fifth of that of Delay-on-Miss. 

We have also evaluated an artificial version of \recomp{} where we keep the same slice coverage but reduce the cost of the slices to at most two cycles. This version exhibits almost identical performance to the real \recomp{}, with a mean performance difference of half a percentage point. This strongly indicates that instead of trying to keep the cost of the slices low, it is more important to increase the coverage, even if large slices are required. This is further corroborated by the results from the Oracle version, discussed below. However, large slices do increase the energy usage, as we will see in Section~\ref{sec:energy}, so a balance still needs to be kept.

If we introduce an Oracle \recomp{} that can recompute all shadowed L1 misses, the difference between the VP and the \recomp{} approaches becomes even more apparent. 
Both Oracle versions have $100\%$ coverage and the same latency, the only difference is that with VP the loads need to be validated when they are unshadowed, while with \recomp{} they are completed as soon as the value has been recomputed.
While, as we have seen, the VP Oracle can only achieve marginal improvements over the non-Oracle version, the \recomp{} Oracle is able to outperform even the baseline, including benchmarks such as \texttt{mcf}, \texttt{cactusADM}, and \texttt{libquantum}.
Of course, such an Oracle is unrealistic, but it does support our argument that the limiting factor for VP is the cost of validation.

However, it is worth noting here that a 100\%-coverage {\recomp} does not necessarily guarantee that the performance will exceed that of the baseline. In fact, there are four benchmarks where the Oracle VRC is slower than the baseline: \texttt{bwaves}, \texttt{milc}, \texttt{leslie3d}, and \texttt{lbm}. Out of these, the \texttt{bwaves} and \texttt{lbm} {\recomp} Oracle is also slower than DoM.
There are various factors that contribute to this result: In \texttt{bwaves} and \texttt{leslie3d} the L1 and the L2 miss ratio (not shown) is increased significantly with the Oracle; in \texttt{milc} the Oracle increases the number of write misses in the L1 (not shown), as well as the average write miss latency (not shown); finally in \texttt{lbm} a combination of many factors contribute to worse cache performance.
The problem is that, even with $100\%$ coverage, not every single memory access is recomputed: Stores, non-speculative loads, and speculative L1 misses that hit in the MSHRs, are still served by the memory hierarchy. By recomputing the rest of the loads, which account for the majority of the L1 misses, the Oracle \recomp{} disrupts the normal operation of the cache and the prefetcher, resulting in performance losses. Essentially, there is a trade-off between the benefits of eliminating long-latency L1 misses and the cost of disrupting the normal cache operation.
For the majority of the benchmarks, this trade-off leans towards the benefits, but this is not true for all of the benchmarks.
Future work aiming to increase {\recomp} coverage must account for these factors to achieve optimal performance.

\subsection{Energy}
\label{sec:energy}

Energy, in our case, is affected by three main factors: The execution time/performance, the number of accesses in the memory hierarchy (especially the DRAM), and the cost of predicting (VP) or recomputing (VRC) a value.
\autoref{fig:energy} shows, starting from the bottom, the dynamic (bottom, light color) and static (middle, dark color) energy of the CPU, the total DRAM energy (middle, light), and, finally, the overhead (if any) for VP and VRC (top, dark). %
Overall Delay-on-Miss and VP increase the mean energy usage over the unsecured baseline by $6\%$, while \recomp{} increases it by $5\%$. The dynamic energy of the CPU (excluding the overheads) remains mostly the same across all versions, instead it is the static, DRAM, and overhead energy that changes.

Static energy is affected because the execution time is affected. This is most obvious in \texttt{mcf}, the application with the worst DoM performance, followed by \texttt{milc}. 
{N}one of the evaluated solutions affect the 
{LLC} MPKI significantly 
{(not shown)}, so the increase in the DRAM energy is not due to an increase in the number of accesses but due to other operations such as refresh and power-down states. These operations do depend on the access patterns, but they also depend on the execution time, similar to the static energy usage of the system.

On the other hand, the overheads introduced by the VP and the \recomp{} are affected both by the execution time (static energy) and by the operations performed. This is particularly visible in the case of the \recomp{}, where the majority of the overhead is due to the instructions of the slices. As we have discussed in Section~\ref{sec:performance}, smaller slices do not lead to better performance, but the same is not true for the energy costs. Instead, a balance between coverage (which increases the performance) and slice length (which increases the energy usage) needs to be achieved. 

Out of all the benchmarks, the ones with the highest (relative to the baseline) energy usage are \texttt{milc} (at $37\%$ over the baseline), \texttt{gromacs} ($13\%$) and \texttt{libquantum} ($12\%$). The rest of the benchmarks have energy overheads of less than $10\%$ over the baseline. \texttt{milc} is the benchmark with the worse performance, so part of the energy increase is due to static and DRAM energy. It also has a high \recomp{} coverage and also some of the third most expensive (in cycles, on average) slices among all the benchmarks, which increases the \recomp{} overhead energy. On the other hand, \texttt{gromacs}'s performance comes very close to the baseline, but it does have the second most expensive slices, while also having high coverage. Finally, \texttt{libquantum} also sees an increase in execution time and by extension, energy usage. The next benchmark with the higher energy increase over the baseline is \texttt{mcf} ($9\%$), but this is far better than DoM, with or without VP, which is at $63\%$ and $84\%$ respectively.



\subsection{
{Hardware/Software Overhead}}
\label{sec:vrcOvhd}

Thus far, related security proposals exert a toll on performance and/or increase cost/complexity. In \arch, as well, microarchitectural support for VRC increases hardware complexity, but only slightly: Slices differ in length, but here we conservatively assume that all would be as long as the maximum-length slice we observe across all benchmarks. In this case, 22 KiB suffices to accommodate all ``live'' slices 
{in Hist, which represents the largest structure}. This is similar to the storage overhead of the VTAGE value predictor we use for the VP configurations. Furthermore,
{static} loads that need to be recomputed 
{at runtime} are few, so the overhead in the binary is small; <3\% across all applications. Finally, as we pointed out throughout the evaluation, since our conservative VRC implementation leaves many optimization opportunities untapped, it still has potential for even further improvement.

\section{Related Work}
\label{sec:rel}
\ignore{
{\color{red} Stefanos: I've got this}
\begin{verbatim}
    1 Making speculation "invisible"
      Can we come up w/ a taxonomy?
        Coverage, complexity, ...?
    2 Cache side channels 
        [Omit?]
    3 Other related work?
        [Omit?]
\end{verbatim}
}

The architecture community promptly proposed a number of techniques (starting with the ground-breaking \emph{InvisiSpec} work~\cite{yan_invisispec:MICRO2018}) to prevent \emph{disclosure gadgets} from revealing secrets. 
The techniques fall in one of the following three broad categories shown below but each individual proposal has different assumptions as to the threat model (type of speculative shadows covered) and prevention of information leakage (disclosure gadgets). 
It is obvious that at this point no direct comparison is possible but we make an effort to compare the solutions qualitatively.

\textbf{Hide\&Replay:} Perform speculative memory accesses in a manner that does not perturb any $\mu$--architectural state in the memory system;
subsequently, perform a replay of the access (when it becomes non-speculative) to affect the correct changes in the $\mu$-architectural state~\cite{li_conditional_2019, khasawneh_safespec:_2019, yan_invisispec:MICRO2018, sakalis+:CF2019ghost, ainsworth+:ISCA2020}. \emph{Invisispec} (Yan et al.)~\cite{yan_invisispec:MICRO2018} and \emph{Ghost loads} (Sakalis et al.)~\cite{sakalis+:CF2019ghost} were the first such proposals. 
Hide\&Replay techniques, as the first to be proposed, showed a significant cost in performance (and a moderate implementation cost). They only protect against information leaks via the memory hierarchy (and not even all of it, as DRAM leaks are possible~\cite{pessl2016drama}). On the other hand, both of these techniques were designed to protect against attacks on any possible speculation primitive, i.e., cover all the speculative shadows mentioned above. 
{
A recent work, InvarSpec~\cite{invarSpec-micro2020} relies on compile time analysis to identify instructions that may become non-speculative during execution (i.e., speculation invariant).
The protection scheme used for these speculative instructions can be lifted at runtime, thus, reducing the performance overhead associated with speculation-related protection mechanisms in hardware. 
Reported performance improvement from such HW-SW co-design, however, cannot reach the negative overhead of \arch. }
{
Instead, as it takes an orthogonal approach, InvarSpec can be used in conjunction with \arch\ to further improve the performance while also reducing the size of the structures needed for recomputation, by reducing the number of loads that trigger recomputation.
}

\textbf{Delay:} Delaying speculative changes in $\mu$-architectural state until execution is non-speculative. Sakalis et al. proposed to delay loads that miss in the L1 (\emph{Delay-on-Miss}) until they are non-speculative~\cite{sakalis+:ISCA2019vp, dom-tc2020}. This delays any $\mu$-state change in the memory hierarchy. A different form of delay 
{(such as \emph{NDA}, proposed by Weisse at al.~\cite{weisse2019nda})}, is to prevent speculative data propagation by delaying \emph{dependent instructions} from executing with speculative inputs~\cite{weisse2019nda, yu_speculative:MICRO2019-STT, fustos+:DAC2019spectreguard, 8891669, schwarz_context:_2019, taram_context-sensitive_2019}. 
Delay-on-Miss protects against \emph{all} speculative shadows (i.e., any possible ``Speculation Primitive'') but delays only changes in the memory hierarchy (including DRAM).
Subsequent work, that delays speculative propagation of data~\cite{weisse2019nda}, achieves good performance by protecting against any $\mu$-state changes (i.e., a much larger gamut of ``disclosure gadgets'' than just the memory hierarchy) but responding only to C-Shadows, i.e., control speculation primitives. {
Another similar alternative, STT~\cite{yu_speculative:MICRO2019-STT}, also protects against other shadows (referred to as the ``Futuristic'' model) but at a higher performance cost. In a recent publication, STT has been extended to utilize speculation as well, referred to as ``speculative data-oblivious speculation -- SDO''~\cite{yu+:ISCA2020}, in order to replace the potentially leaky speculative paths with secure, data-independent, paths. This approach is similar to the approach that ISER takes, only ISER is non-speculative and does not require any verification or squashing, further reducing the runtime overhead.
{
Tran et al., propose a SW-HW extension that can
reduce the time in which loads are shadowed (i.e., loads are speculative) and 
thereby can increase the MLP~\cite{clearing-shadows-pact2020}. Their proposal includes instruction reordering to prioritize calculations that 
minimize the speculation window, such as target address computation of memory accesses, and resolution of branch conditions. Much like InvarSpec, their approach may reduce the performance overhead of delay-based security solutions by reducing the number of speculative loads or time spent in speculations, and it is orthogonal to our proposal.
}
Both approaches can be combined together to offer better security coverage with minimum performance overhead.}
{
SPECCFI~\cite{speccfi20} uses the Control-Flow Integrity (CFI) to prevent Spectre-type attacks that abuse illegal control flow during speculative execution. Not all possible speculative side-channel attacks are covered by this technique but, much like the other compiler-based techniques we have discussed, it can be used in conjunction with our technique to 
limit the cases where recomputation is needed. 
}

\textbf{Cleanup:} Perform a speculative change in $\mu$-architectural state but then \emph{undo} if speculation is squashed. In the first such proposal, \emph{CleanupSpec}, by Sailshwar et al.~\cite{saileshwar2019cleanupspec}, the undo is expensive so its application is restricted to the L1 cache. The rest of the memory hierarchy (L2, LLC, and coherence Directory) is assumed to be protected in other ways, including randomization and delaying of coherence state changes, but DRAM row buffers still remain a security hole.
Cleanup techniques only protect the L1, assuming---at a cost---that the rest of the hierarchy (excluding DRAM) is protected otherwise~\cite{saileshwar2019cleanupspec}.

\textbf{(Generic) Recomputation:} Amnesiac~\cite{amnesiac17} introduces a $\mu$-architecture for recomputation that differs from \arch{} in the way slices are generated and their usage.
The goal of Amnesiac is to replace as many energy-hungry loads as possible with recomputations of the respective data value. 
In contrast, ISER recomputes slices selectively, such that recomputation is triggered only for shadowed loads that miss in L1.

Kandemir et al., proposed a recomputation-based approach to reduce off-chip memory space in embedded processors~\cite{Kandemir:2005gw}.  
Koc et al. investigated how recomputation of data residing in memory banks in low-power states can reduce the energy consumption~\cite{Koc:2006ce}, and devised compiler optimizations for scratchpads~\cite{Koc:2007bl} that are limited to array variables.
The dual of recomputation, \emph{memoization}~\cite{Sodani:1997hn, resistiveComp} replaces computation with table look-ups for pre-computed values (for the ones that are frequent and expensive to recompute). Memoization can mitigate the communication overhead -- as long as  table
look-ups are cheaper than long-distance data retrieval, but 
is only effective if the respective computations exhibit significant value locality. 
Therefore, memoization and recomputation can complement each other in boosting energy efficiency.

\emph{Idempotent Processors}~\cite{idem} execute programs as a sequence of
compiler-constructed idempotent (i.e., re-executable without any side effects) code regions.  
As the name suggests, idempotent regions regenerate the same output regardless of how many times they are executed with the given program state.  
Generally, idempotent regions are larger, and therefore tend to incur higher overhead for recomputation, while slices for VRC employ fine-grain data recomputation (along a short, independent slice for each value), where each slice contains only the necessary instructions to generate a value. 
Accordingly, slices for VRC may provide more flexibility than idempotent regions.

Elnawawy et al., demonstrated the applicability of recomputation to loop-based
code~\cite{Elnawawy2017} to reduce checkpointing overheads. 
In their proposal, a whole
loop is (re)executed during recovery, where only the initial state of the loop is
required to be checkpointed. 
The loops may contain extra computations that
are not relevant to the production of the value to be recovered. 
Compared to such a coarse-grain recomputation,
slice-based recomputation does not contain any irrelevant instructions. Also, slices used for VRC do not contain load
instructions, as opposed to~\cite{Elnawawy2017}; and recomputation applies outside of loops, providing wider applicability.
To summarize, although value recomputation has been explored in different contexts before, to the best of our knowledge, none of the prior works has evaluated recomputation in the context of security.

{\bf Slice Generation:} 
Automatic creation of VRC slices in hardware is complicated because we are not after the slice of the load to be replaced (which could be created by existing techniques like IBDA~\cite{ibdaIsca15}) but the slice of the corresponding store creating the value. This would require tracking of all stores and their slices and somehow matching these with a (speculative) load missing in the L1 cache.
Srinivasan et al.~\cite{contFlowPipe} generate ``forward'' slices for loads that miss in LLC. This is easier on hardware since the dependency tracking starts with the producer (i.e., load that misses in LLC) and the consumers (following use-def chains) are executed after the producer. However, in our case, we have to identify ``backward'' slices -- i.e., the producers, not consumers, of a value that will be loaded -- where all the producers were executed before the load itself. Such backward dependency tracking would likely require expensive bookkeeping in hardware.%

\section{Conclusion}
\label{sec:conc}
\ignore{
\begin{verbatim}
    Summary
        Problem, solution, evaluation
    Discuss 
        Coverage
        Limitations
        New vulnerabilities? 
\end{verbatim}
}
\emph{Delay} techniques aim to hide the effects of transient execution by simply delaying instructions until they become non-speculative. 
Whether delaying loads that miss in the L1, as Delay-on-miss does, or delaying the propagation of speculative data to dependent instructions, as NDA and STT do, delay techniques extract a heavy toll in performance, in direct relation to the set of speculative shadows they protect against. 
Delay techniques would be at an impasse with respect to improvement if we could not regain some of this lost performance in some other way. To this end, value prediction, invisible from the outside, was initially proposed as a solution.

However, value prediction is not the right abstraction for recovering lost performance in Delay-on-miss. This is not because of coverage or accuracy but because value prediction is just another form of speculation that needs to be validated. Validation limits the potential benefits to the point where even an oracle VP (100\% coverage and accuracy) does not do any better than a practical VP. In our evaluation we found that, no matter how good, VP is limited to just one percentage point improvement over Delay-on-miss.

Instead, we propose another, \emph{non-speculative}, abstraction to regain performance for delay techniques, and in particular for Delay-on-miss. We propose to use recomputation that yields correct values---not predictions---as the key to overcome Delay-on-miss performance limitations. We describe the architecture, we evaluate it using a practical approach to generate recomputation slices albeit with modest coverage, and we exceed the performance of Oracle VP ($90\%$ vs. $93\%$) with lower energy usage. Finally, we discuss the potential for increasing the coverage of recomputation with future architectural support. Because, as we show, oracle recomputation easily exceeds even the performance of the unmodified (unsecured) baseline, this direction provides tangible motivation for researching techniques for a future secure processor. 

To regain the performance cost of securing the memory hierarchy we need to identify methods that improve the MLP. This paper demonstrates, for the first time, value recomputation's unique ability in overcoming the MLP restriction that is inherent in VP when applied on the Delay-on-Miss technique. To the best of our knowledge, {\em no previous study on recomputation considered any security impact}. 
%
{Finally, these findings should be considered in the context of our representative threat model (Section}~\ref{sec:threat}). 
In the end, no threat model can cover all possible security vulnerabilities. 
But, as explained in Section~\ref{sec:recmp-sec}
{, \arch{} 
does not introduce any new attack opportunities under the provided threat model.

}

That said, as any technique that affects the control flow timing -- including value prediction or even Delay-on-Miss to name a few -- recomputation may give rise to timing channels, where information to be leaked gets encoded in timing differences between various microarchitectural events. 
Even if a given value is recomputed multiple times throughout execution, as resource contention and  
speculation can easily change timing of microarchitectural events non-deterministically, there is also a  very good chance that recomputation rather {\em obfuscates} control flow timing. 

Specifically, provided that 
(i) a slice is executed only upon an associated L1 miss; that
(ii) each slice may have not only a different number but also a different composition of arithmetic/logic instructions, with each instruction featuring a different number of operands (which all affect how much time it takes to process each slice instruction through the SFile); that (iii) none, one, or more Hist accesses may be the case per slice execution; and that (iv) due to the small footprint of such microarchitectural buffers, their access times are relatively short, identifying a unique timing signature for each slice encountered throughout the execution (in order to associate slice timings with values) would not be easy -- even more so under speculation and resource contention.

To conclude,  potential timing channels, if at all, would not necessarily be straight-forward to exploit. In fact,  recomputation is more likely to result in control flow obfuscation.  We leave the exploration of such  effects to future work, confining the analysis in this paper to only memory side-channels because they are  easier to exploit and can be exploited across cores. This does not imply that side-channels such as functional unit contention based ones are not possible, they are just outside the scope of our threat model. 

\bibliographystyle{unsrt}  

\bibliography{refs}

\end{document}